\newcommand{\vA}{\boldsymbol{A}}
\newcommand{\vf}{\boldsymbol{f}}
\newcommand{\vfDL}{\boldsymbol{f}_{\text{DL}}}
\newcommand{\vg}{\boldsymbol{g}}

\newcommand{\vp}{\boldsymbol{p}}
\newcommand{\vP}{\boldsymbol{P}}
\newcommand{\wTV}{||\boldsymbol{f}^{(n)}||_{\text{wTV}}}
\newcommand{\subjt}{\text{ subject to }}

\documentclass[journal]{IEEEtran}
\usepackage{graphicx}
\usepackage[tight]{subfigure}
\usepackage{color}

\usepackage{cite}

\ifCLASSINFOpdf
\else
\fi
\usepackage{amsmath}
\usepackage{algorithmic}
\ifCLASSOPTIONcompsoc
  \usepackage[caption=false,font=normalsize,labelfont=sf,textfont=sf]{subfig}
\else
  \usepackage[caption=false,font=footnotesize]{subfig}
\fi
\begin{document}
%
\title{Data Consistent CT Reconstruction from Insufficient Data with Learned Prior Images}
%
%
%

\author{Yixing~Huang,
        Alexander~Preuhs,
        Michael~Manhart,
        Guenter~Lauritsch,
        Andreas~Maier
\thanks{Y.~Huang, A.~Preuhs and A.~Maier are with Pattern Recognition Lab, Friedrich-Alexander-University Erlangen-Nuremberg, Erlangen, Germany (e-mail: yixing.yh.huang@fau.de).}
\thanks{A.~Maier is also with Erlangen Graduate School in Advanced Optical Technologies (SAOT), Erlangen, Germany.}
\thanks{M.~Manhart and G. Lauritsch are with Siemens Healthcare, GmbH, Forchheim, Germany.}
}

%
%

\markboth{ }%
{Huang \MakeLowercase{\textit{et al.}}: Data Consistent CT Reconstruction}
%



\maketitle

\begin{abstract}
Image reconstruction from insufficient data is common in computed tomography (CT), e.\,g., image reconstruction from truncated data, limited-angle data and sparse-view data. Deep learning has achieved impressive results in this field. However, the robustness of deep learning methods is still a concern for clinical applications due to the following two challenges: a) With limited access to sufficient training data, a learned deep learning model may not generalize well to unseen data; b) Deep learning models are sensitive to noise. Therefore, the quality of images processed by neural networks only may be inadequate. In this work, we investigate the robustness of deep learning in CT image reconstruction by showing false negative and false positive lesion cases. 
Since learning-based images with incorrect structures are likely not consistent with measured projection data, we propose a data consistent reconstruction (DCR) method to improve their image quality, which combines the advantages of compressed sensing and deep learning: First, a prior image is generated by deep learning. Afterwards, unmeasured projection data are inpainted by forward projection of the prior image. Finally, iterative reconstruction with reweighted total variation regularization is applied, integrating data consistency for measured data and learned prior information for missing data. The efficacy of the proposed method is demonstrated in cone-beam CT with truncated data, limited-angle data and sparse-view data, respectively. For example, for truncated data, DCR achieves a mean root-mean-square error of 24\,HU and a mean structure similarity index of 0.999 inside the field-of-view for different patients in the noisy case, while the state-of-the-art U-Net method achieves 55\,HU and 0.995 respectively for these two metrics.
\end{abstract}

\begin{IEEEkeywords}
Deep learning, robustness, compressed sensing, data consistency, insufficient data, computed tomography.
\end{IEEEkeywords}

%
\IEEEpeerreviewmaketitle

\section{Introduction}
%
%
%
%
\IEEEPARstart{C}{omputed} tomography (CT) is a widely used medical imaging technology for disease diagnosis and interventional surgeries. CT reconstructs a volume which provides cross sectional images and offers good visualization of patients' anatomical structures and disease information. In order to acquire sufficient data for image reconstruction, certain acquisition conditions need to be satisfied. First of all, the detector of a CT system needs to be large enough to cover an imaged object. Second, the angular range of a scan should be at least $180^\circ$ plus a fan angle. Third, the angular step should be small enough to meet sampling theorems \cite{natterer2001mathematics}. 

However, such conditions may be violated in practical applications, raising the issues of interior tomography, limited-angle tomography, and sparse-view reconstruction. In interior tomography, X-rays are collimated to a certain region-of-interest (ROI) to reduce the amount of dose exposure to patients. In addition, data truncation is a common problem for large patients whose body cannot be entirely positioned inside the field-of-view (FOV) due to the limited detector size. The problem of limited-angle tomography arises when the rotation of a gantry is restricted by other system parts or an external obstacle. Sparse-view reconstruction is preferred for the sake of low dose, quick scanning time, or avoidance of severe motion. In these situations, artifacts, typically cupping artifacts, streak artifacts and view aliasing, occur due to missing data.


To deal with missing data, data inpainting is the most straight-forward solution. For interior tomography, heuristic extrapolation methods are widely applied, including symmetric mirroring \cite{ohnesorge2000efficient}, cosine function fitting \cite{sourbelle2005reconstruction}, and water cylinder extrapolation (WCE) \cite{hsieh2004novel}. With such extrapolations, a smooth transition between measured and truncated areas is pursued to alleviate cupping artifacts inside the FOV. However, anatomical structures outside the FOV are still corrupted. For limited-angle and sparse-view reconstruction, many researchers attempted to restore missing data based on sinusoid-like curves \cite{li2012strategy,kim2018feasibility}, band-limitation properties \cite{defrise1983regularized,qu2009landweber,pohlmann2014estimation,huang2018papoulis}, and data consistency conditions \cite{louis1980picture,prince1990constrained,kudo1991sinogram,huang2017restoration}. Such approaches achieved improved image quality for certain scanning configurations and particular subjects only, but only achieve limited performance for clinical applications. 

With the advent of compressed sensing technologies, iterative reconstruction with total variation (TV) regularization became popular for CT reconstruction from insufficient data. So far, many TV algorithms have been developed, including ASD-POCS \cite{sidky2008image}, improved TV (iTV) \cite{ritschl2011improved}, spatio-temporal TV (STTV) \cite{taubmann2017spatio}, anisotropic TV (aTV) \cite{chen2013limited}, soft-thresholding TV \cite{yu2010soft}, total generalized variation (TGV)\cite{niu2014sparse}, and scale-space anisotropic TV (ssaTV) \cite{huang2018scale}. For insufficient data, they achieve superior image quality compared to FBP-based reconstruction, as TV regularization takes the advantage of the sparsity prior in the gradient domain. Except for image gradient domain, sparsity prior can also be employed in other transformed domains \cite{zhu2013improved,frikel2013sparse,li2014dictionary}.

Recently, deep learning has achieved impressive results in CT reconstruction \cite{wang2016perspective,wang2018image}. For image reconstruction from insufficient data, deep learning has been applied for sinogram inpainting in the projection domain \cite{lee2017view,ghani2018deep,lee2018deep,anirudh2018lose,li2019promising}, artifact post-processing in the image domain and image transformed domains \cite{wolterink2017generative,gu2017multi,xie2018artifact,dai2018limited,schwab2019deep,yang2018low,han2018framing,han2019one,liao2018adversarial,fournie2019ct}, and direct projection-to-image reconstruction \cite{zhu2018image,fu2019hierarchical,chen2018learn,wurfl2018deep,adler2018learned,li2019learning,vishnevskiy2019deep}. For sinogram inpainting and artifact post-processing, incomplete sinograms and artifact corrupted images need to be translated to complete sinograms and artifact-free images, respectively. For these tasks, the U-Net \cite{ronneberger2015u} and generative adversarial networks (GANs) \cite{goodfellow2014generative,yi2019generative} are the most frequently used techniques. For projection-to-image reconstruction, known operators for image reconstruction \cite{maier2019learning} are typically integrated into the architecture design of deep neural networks \cite{wurfl2018deep,adler2018learned,li2019learning,vishnevskiy2019deep}.

The above achievements have shown a promising prospect of the clinical application of deep learning into CT reconstruction. However, the robustness of deep learning in practice is still a concern \cite{fawzi2017robustness,liu2018towards,yuan2019adversarial}. It is well-known that deep learning cannot generalize well to unseen data \cite{zhang2016understanding}, especially with insufficient training data. Meanwhile, it is reported that adversarial examples \cite{szegedy2013intriguing,goodfellow2014explaining,kurakin2016adversarial,tang2019adversarial} are ubiquitous in deep neural networks. While small perturbations hardly visible to human eyes can cause deep neural networks to predict an entirely wrong label \cite{szegedy2013intriguing,goodfellow2014explaining,kurakin2016adversarial}, objects with significant changes might be classified as the same label by deep neural networks since a trained neural network typically cannot capture all necessary features to distinguish objects of different classes \cite{tang2019adversarial,wang2020cross}. 
The instability of deep learning is widely investigated in the field of computer vision \cite{fawzi2017robustness,liu2018towards,yuan2019adversarial,zhang2016understanding,szegedy2013intriguing,goodfellow2014explaining,kurakin2016adversarial,tang2019adversarial}. However, it has not been adequately investigated in the field of CT reconstruction. In our previous work \cite{huang2018some}, we found that deep learning is sensitive to noise in the application of limited-angle tomography. In this work, more false negative and false positive examples will be given for the applications of deep learning in CT reconstruction from insufficient data.

Due to the factors of insufficient training data and noise as aforementioned, generating reconstructed images directly from a neural network appears inadequate since incorrect structures might occur in the learned images. Learned images with incorrect structures are likely not consistent with measured projection data. Therefore, enforcing data consistency can improve their image quality in principle. 
Since images generated by deep learning methods can provide substantial beneficial information of anatomical structures, they potentially offer good prior for unmeasured projection data. Accordingly, we propose a data consistent reconstruction (DCR) method using learned prior images for CT reconstruction from insufficient data, where iterative reconstruction with TV regularization is applied to integrate data consistency for the measured data and learned image prior for the unmeasured data. 

This work is an extension of our previous preliminary work \cite{huang2019data} and \cite{huang2020field}. The major contributions of this work lie in the following two aspects: a) Investigate the robustness of deep learning in CT image reconstruction by showing false negative and false positive lesion cases;  b) Propose the DCR method to improve the image quality of learned images for CT reconstruction from insufficient data, which is a hybrid method to combine the advantages of deep learning and compressed sensing.

\section{Materials And Methods}
Our proposed DCR method mainly consists of three steps: artifact reduction using deep learning, data inpainting with learned prior images, and iterative reconstruction with TV regularization.

\subsection{Artifact Reduction Using Deep Learning}
\begin{figure}[h]
\centering
\includegraphics[width = 1\linewidth]{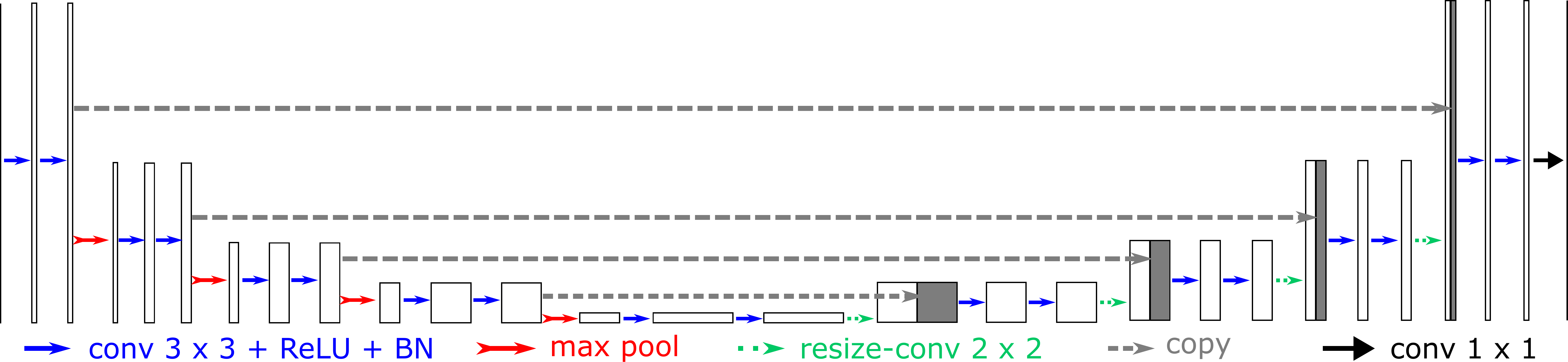}
\caption{The U-Net architecture for artifact reduction in images reconstructed from insufficient data.}
\label{Fig:UNetArch}
\end{figure}

\subsubsection{Neural network}
As mentioned above, various neural networks have been reported effective for image reconstruction from insufficient data, for example, the U-Net, GANs, or the iCT-Net \cite{li2019learning}. In principle, images processed by any one of these networks can provide prior information for the inpainting of missing data. In this work, we choose the state-of-the-art U-Net for its concise architecture design, fast convergence in training and high representation power. Its architecture is displayed in Fig.\,\ref{Fig:UNetArch}.  

\subsubsection{Input}
For limited-angle tomography and sparse-view CT, images reconstructed by FBP directly from insufficient data, denoted by $\vf_{\text{FBP}}$, are used as input images.
For truncation correction, it is more effective for a neural network to learn artifacts in images reconstructed by FBP from extrapolated data than truncated data directly, according to our previous research \cite{fournie2019ct,huang2020field}.
Therefore, in this work, an image reconstructed from WCE \cite{hsieh2004novel} processed projections, denoted by $\vf_{\text{WCE}}$, is chosen as the input to the network. 

As the artifacts are mainly caused by erroneous extrapolation of WCE, sparse-view sampling or limited-angle acquisition, the influence of cone-beam artifacts is neglected. To reduce computational burden, we process the data slice-wise instead of volume-vise. The correlation among different 2-D slices in the third dimension will be brought back in the third step of DCR. 

\subsubsection{Output}
The training target of the network is the corresponding artifact image of the input image, denoted by $\vf_{\text{artifact}}$. It is computed as the subtraction of the input image by the data-complete reconstruction. For training, the $\ell_2$ loss function is used for truncation correction and limited-angle tomography, while the perceptual loss in \cite{stimpel2017mr} is used for sparse-view CT since the $\ell_2$ loss is not sensitive to high frequency, low magnitude aliasing. For inference, the network estimates the artifact image $\vf'_{\text{artifact}}$, and an artifact-free image $\vfDL$ is computed by the subtraction of the artifact image from the input image, i.\,e., $\vfDL = \vf_{\text{WCE}} - \vf'_{\text{artifact}}$ for truncation correction and $\vfDL = \vf_{\text{FBP}} - \vf'_{\text{artifact}}$ for sparse-view and limited-angle reconstruction.

\subsection{Data Inpainting with Learned Prior Images}

For data consistent reconstruction, we propose to preserve measured projections entirely and use the learned reconstruction for data inpainting of missing projections. We denote measured projections by $\vp_{\text{m}}$ and their corresponding system matrix by $\vA_{\text{m}}$. We further denote unmeasured projections by $\vp_{\text{u}}$ and their corresponding system matrix by $\vA_{\text{u}}$. The learned reconstruction $\vfDL$ provides prior information for the unmeasured projections $\vp_{\text{u}}$. Therefore, an estimation of $\vp_{\text{u}}$, denoted by $\hat{\vp}_{\text{u}}$, is achieved using digitally rendered radiographs (DRRs) of $\vfDL$,
\begin{equation}
\hat{\boldsymbol{p}}_{\text{u}} = \boldsymbol{A}_{\text{u}} \vfDL.
\end{equation} 
Combining $\hat{\boldsymbol{p}}_{\text{u}}$ with $\boldsymbol{p}_{\text{m}}$, a complete projection set is obtained for reconstruction.

\subsection{Iterative Reconstruction with TV Regularization}

Due to intensity discontinuity between $\hat{\boldsymbol{p}}_{\text{u}}$ and $\boldsymbol{p}_{\text{m}}$ at the transition area as well as noise in the measured projections, iterative reconstruction with TV regularization is applied to further reduce artifacts and noise. 

In this work, particularly the iterative reweighted total variation (wTV) regularization \cite{huang2018scale} is utilized, as wTV intrinsically reduces staircasing effect and preserves fine structures well. The wTV of an image $\vf$ is defined as \cite{candes2008enhancing},
\begin{equation}
\begin{split}
&||\boldsymbol{f}^{(n)}||_{\text{wTV}}=\sum_{x,y,z}\boldsymbol{w}^{(n)}_{x,y,z}||\mathcal{D}\boldsymbol{f}^{(n)}_{x,y,z}||,\\
&\boldsymbol{w}^{(n)}_{x,y,z}=\frac{1}{||\mathcal{D}\boldsymbol{f}^{(n-1)}_{x,y,z}||+\epsilon},
\end{split}
\label{eq:WeightsUpdate}
\end{equation}
where $\boldsymbol{f}^{(n)}$ is the image at the $n^\text{th}$ iteration, $\boldsymbol{w}^{(n)}$ is the weight vector for the $n^\text{th}$ iteration which is computed from the previous iteration, and $\epsilon$ is a small positive value added to avoid division by zero. A smaller value of $\epsilon$ results in finer image resolution but slower convergence speed.

The overall objective function for the $n^\text{th}$ iteration is,
 \begin{equation}
 \min\wTV, \subjt \left\lbrace
 \begin{array}{l}
 ||\vA_{\text{m}}\vf - \vp_{\text{m}}|| < e_1,\\
 ||\vA_{\text{u}}\vf - \hat{\vp}_{\text{u}}|| < e_2,
 \end{array} 
 \right.
 \label{eqn:ObjectiveDataConsistentDeepLearning}
 \end{equation}
with initialization,
\begin{equation}
\vf^{(0)} = \vfDL.
\end{equation}
Here $e_1$ is a noise tolerance parameter for the data fidelity term of the measured projections and $e_2$ is a tolerance parameter that accounts for the inaccuracy of the prior image $\vfDL$. The iterative reconstruction is initialized by $\vfDL$ to accelerate convergence. With the above objective function, the data consistency for the measured data and the learned image prior for the unmeasured data are integrated.

 To solve the above objective function, simultaneous algebraic reconstruction technique (SART) is utilized to minimize the data fidelity constraints, while a gradient descent method is utilized to minimize the wTV term \cite{huang2018scale}. The implementation details of the SART + wTV algorithm  are presented in the following pseudo-code. The error tolerance parameters $e_1$ and $e_2$ are integrated by two soft-thresholding operators $\mathcal{S}_{e_1}$ and $\mathcal{S}_{e_2}$ respectively in Line 9-13, where 
 \begin{equation}
 \mathcal{S}_{\tau}(x)=\left\lbrace
 \begin{array}{ll}
 x - \tau, & x > \tau,\\
 0, & |x|\leq \tau, \\
 x + \tau, &x < -\tau.
 \end{array} 
 \right.
 \end{equation}

\begin{algorithmic}[1]
 \STATE{Initialization: $\vfDL$, $e_1$, $e_2$, $\epsilon$, $\lambda$, $n_{\max}$, $n=0$}
 \STATE{$\vf^{(0)} = \vfDL$}
 \STATE{$\boldsymbol{w}^{(0)}_{x,y,z}={1}/{\left(||\mathcal{D}\boldsymbol{f}^{(0)}_{x,y,z}||+\epsilon\right)}$}
 
 \REPEAT 
 	\STATE{SART update for data fidelity:}
 	\FOR{ $\beta = \beta_{\min}; \beta <= \beta_{\max}; \beta = \beta + \Delta\beta$}
 	 	
 		\FORALL{$\vp_i \in \vP_{\beta}$ simultaneously}
  			\STATE{$\vp^{(n)}_i :=\sum^N_{k=1}\vA_{i,k}\cdot \vf_k^{(n)}$}
  			\IF {$\vp_i$ is measured}
				\STATE{$\Delta \vp^{(n)}_i := \mathcal{S}_{e_1} (\vp_i - \vp^{(n)}_i)$}
  			\ELSE
				\STATE{$\Delta \vp^{(n)}_i := \mathcal{S}_{e_2} (\hat{\vp}_i - \vp^{(n)}_i)$}  
  			\ENDIF
  			\STATE{$\Delta \vp^{(n)}_i := \Delta \vp^{(n)}_i/{\left(\sum_{k=1}^{N}\vA_{i,k} \right)}$}
  		\ENDFOR

  		\FORALL{$\vf_j\in \vf$ simultaneously} 
  			\STATE{$\Delta \vf_j :=  \frac{\sum_{\vp_i\in \vP_{\beta}}{\Delta \vp^{(n)}_i}\cdot \vA_{i,j}}{\sum_{\vp_i\in \vP_{\beta}}\vA_{i,j}}$}
  		\ENDFOR
  
  		\STATE{$\vf^{(n)} := \vf^{(n)}+\lambda \cdot \Delta \vf$}

 	\ENDFOR
 	
 	\STATE{$\ $}
 	\STATE{Enforce nonnegativity:} 
 	\FORALL{$\vf_j\in \vf$}		
 		\STATE{$\vf_j = (\vf_j <0)\ ?\ 0 : \vf_j$}	
 	\ENDFOR
 	
 	\STATE{$\ $}
	\STATE{wTV minimization:} 	
 		\FOR{$l = 0; l < l_{\max}; l$++} 
 	 		\STATE{compute wTV gradient:}
 	 		\STATE{$\vg := \nabla{\wTV}$}
 	 		\STATE{$\vg := \vg/||\vg||_{\infty}$}
 	 		\STATE{backtrack line search for $t$:}
 	 		\STATE{$t := 1.0, \alpha = 0.3, \gamma = 0.6$}
 	 		\WHILE{$||\vf^{(n)}-t\cdot \vg||_{\text{wTV}} > ||\vf^{(n)}||_{\text{wTV}} + \alpha \cdot t \cdot \vg^\top \vg$}
 	 			\STATE{$t:= \gamma \cdot t$}
 	 			\ENDWHILE
 	 		\STATE{$\vf^{(n)} := \vf^{(n)} - t \cdot \vg$}
 		\ENDFOR
 	\STATE{$\boldsymbol{w}^{(n)}_{x,y,z}={1}/{\left(||\mathcal{D}\boldsymbol{f}^{(n)}_{x,y,z}||+\epsilon\right)}$}
 	\STATE{$n$++}
  \UNTIL{$n = n_{\max}$}
\end{algorithmic}
\subsection{Experimental Setup}
In this work, the proposed DCR method is evaluated on image reconstruction tasks from truncation data, limited-angle data and sparse-view data respectively in cone-beam CT systems.

\subsubsection{System configurations}
 The source-to-detector distance of the cone-beam CT is 1200\,mm and the source-to-isocenter distance is 600\,mm. The reconstruction volume size is $256 \times 256 \times 256$ with a voxel size of 1.25\,mm $\times$ 1.25\,mm $\times$ 1.0\,mm. The detector pixel size is 1.0\,mm $\times$ 1.0\,mm. For a regular short scan, the angular range is $210^\circ$ with an angular step of $1^\circ$ using a large detector size of $1240 \times 960$ pixels. For lateral data truncation study, the detector is switched to a small size of $600 \times 960$ pixels. For sparse-view CT, the angular step is switched to $4^\circ$ in a full scan of $360^\circ$. For limited-angle tomography, the angular range is switch to $150^\circ$. To generate DRRs for data inpainting, a ray-driven forward projection method with a sampling rate of 7.5/mm is applied. For all experiments considering noise, Poisson noise is simulated assuming an exposure of $I_0 = 10^5$ photons at each detector pixel before attenuation.

\subsubsection{Neural network training and test}
In this work, 18 patients' CT data sets \cite{mccollough2017low} are used, with a split of 16-1-1 for training, validation and test, respectively. To investigate the performance of deep learning on different training and test data, leave-one-out cross-validation is performed among 17 patients' data sets, while one patient is always used for validation. For training, 25 slices from each patient are chosen with a neighbouring slice distance of 10\,mm. For validation, the root-mean-square error (RMSE) of 25 slices is used for monitoring the training process to avoid over-fitting. For noise-free evaluation, the U-Net is trained on noise-free data, while it is trained on data with Poisson noise in the noisy cases. The $\ell_2$-norm is applied to regularize neural network weights with a regularization parameter of $10^{-4}$. The U-Net is trained on the above data for 500 epochs. The initial learning rate is $10^{-3}$ and the decay rate is 0.97 for each epoch. For the test patient, all the 256 slices are fed to the U-Net for evaluation.

\subsubsection{Iterative reconstruction parameters}
For iterative reconstruction, the error tolerance parameter $e_1$ is set to 0.005, while a relatively large value of 0.5 is chosen empirically for $e_2$ in Eqn.\,(\ref{eqn:ObjectiveDataConsistentDeepLearning}) in the noise-free case. In the noisy case, $e_1$ is set to 0.05 to incorporate Poisson noise, while $e_2$ is kept as 0.5. Please see Fig.\,\ref{Fig:paraSelection} for the above parameter selection. For the wTV regularization, the parameter $\epsilon$ is set to 5\,HU for weight update in Eqn.\,(\ref{eq:WeightsUpdate}). 10 iterations of SART + wTV are applied using the U-Net reconstruction $\vf_{\text{DL}}$ as initialization to get the final reconstruction, i.e., $n_{\max} = 10$ in the pseudo-code of the SART + wTV algorithm. For SART, the relaxation parameter $\lambda$ is set to 0.8. For wTV minimization, 10 subiterations are applied, i.e., $l_{\max} = 10$. 

\subsubsection{Image quality assessment}
To assess image quality, the metrics of RMSE and structure similarity (SSIM) are utilized. In addition, whether false negative and false positive lesions are reconstructed is analysed, since the standard image quality metrics (e.g., RMSE and SSIM) cannot fully indicate the clinical value of images.

\section{Results}

\subsection{Truncation Correction}
\begin{figure*}[th]
\centering
\begin{minipage}{0.16\linewidth}
\centerline{$\boldsymbol{f}_{\text{reference}}$}
\smallskip
\end{minipage}
\begin{minipage}{0.16\linewidth}
\centerline{$\boldsymbol{f}_{\text{FBP}}$}
\smallskip
\end{minipage}
\begin{minipage}{0.16\linewidth}
\centerline{$\boldsymbol{f}_{\text{WCE}}$}
\smallskip
\end{minipage}
\begin{minipage}{0.16\linewidth}
\centerline{$\boldsymbol{f}_{\text{wTV}}$}
\smallskip
\end{minipage}
\begin{minipage}{0.16\linewidth}
\centerline{$\boldsymbol{f}_{\text{U-Net}}$}
\smallskip
\end{minipage}
\begin{minipage}{0.16\linewidth}
\centerline{$\boldsymbol{f}_{\text{DCR}}$}
\smallskip
\end{minipage}

\begin{minipage}{0.16\linewidth}
\subfigure[]
{
\includegraphics[width = \linewidth]{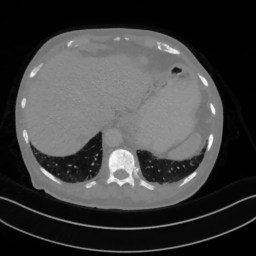}
}
\end{minipage}
\begin{minipage}{0.16\linewidth}
\subfigure[408\,HU]
{
\includegraphics[width = \linewidth]{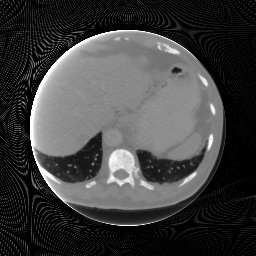}
}
\end{minipage}
\begin{minipage}{0.16\linewidth}
\subfigure[178\,HU]
{
\includegraphics[width = \linewidth]{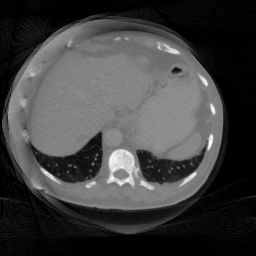}
}
\end{minipage}
\begin{minipage}{0.16\linewidth}
\subfigure[141\,HU]
{
\includegraphics[width = \linewidth]{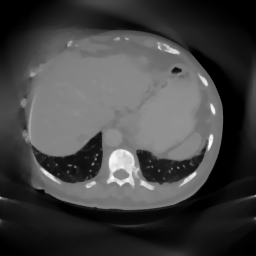}
}
\end{minipage}
\begin{minipage}{0.16\linewidth}
\subfigure[67\,HU]
{
\includegraphics[width = \linewidth]{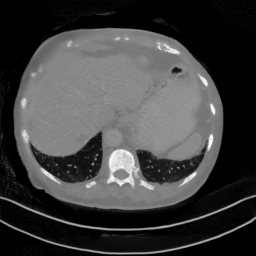}
}
\end{minipage}
\begin{minipage}{0.16\linewidth}
\subfigure[49\,HU]
{
\includegraphics[width = \linewidth]{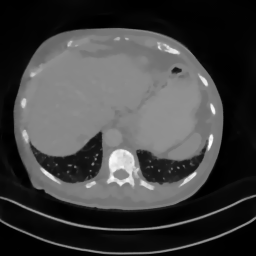}
}
\end{minipage}

\begin{minipage}{0.16\linewidth}
\subfigure[]
{
\includegraphics[width = \linewidth]{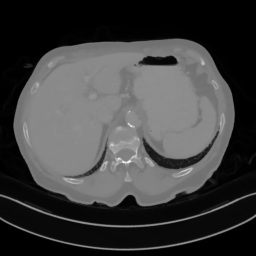}
}
\end{minipage}
\begin{minipage}{0.16\linewidth}
\subfigure[369\,HU]
{
\includegraphics[width = \linewidth]{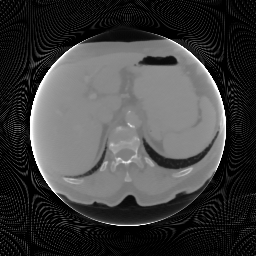}
}
\end{minipage}
\begin{minipage}{0.16\linewidth}
\subfigure[175\,HU]
{
\includegraphics[width = \linewidth]{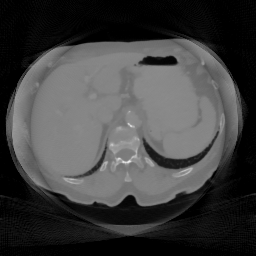}
}
\end{minipage}
\begin{minipage}{0.16\linewidth}
\subfigure[122\,HU]
{
\includegraphics[width = \linewidth]{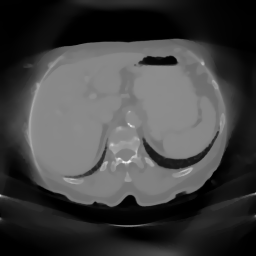}
}
\end{minipage}
\begin{minipage}{0.16\linewidth}
\subfigure[66\,HU]
{
\includegraphics[width = \linewidth]{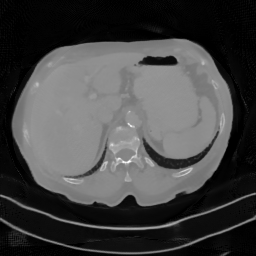}
}
\end{minipage}
\begin{minipage}{0.16\linewidth}
\subfigure[42\,HU]
{
\includegraphics[width = \linewidth]{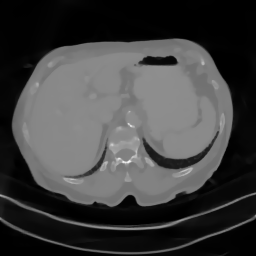}
}
\end{minipage}

\caption{Results of two example slices from the test patients in the noise-free case, window: [-1000, 1000]\,HU. The top and bottom rows are for the 175th slice of Patient NO.\,3 and the 150th slice of Patient NO.\,17, respectively. The RMSE value of the whole body for each method is displayed at the corresponding subcaption.}
\label{Fig:TruncationNoiseFreeLW}
\end{figure*}

The reconstruction results of two example slices from the test patients in the noise-free case are displayed in Fig.\,\ref{Fig:TruncationNoiseFreeLW}. The RMSE of the whole body area for each method is displayed in the corresponding subcaption.
In the FBP reconstruction $\vf_{\text{FBP}}$ (Figs.\,\ref{Fig:TruncationNoiseFreeLW}(b) and (h)), the original FOV boundary is clearly observed. The structures inside the FOV suffer from severe cupping artifacts, while the anatomical structures outside this FOV are entirely missing. 
For $\vf_{\text{WCE}}$ in Figs.\,\ref{Fig:TruncationNoiseFreeLW}(c) and (i), the cupping artifacts inside the FOV are notably alleviated by WCE. Anatomical structures outside the FOV are partially reconstructed. 
In the wTV reconstruction $\vf_{\text{wTV}}$ (Figs.\,\ref{Fig:TruncationNoiseFreeLW}(d) and (j)), the cupping artifacts are removed. Nevertheless, the structures outside the FOV are still missing due to data truncation.
Figs.\,\ref{Fig:TruncationNoiseFreeLW}(e) and (k) demonstrate that the U-Net is able to reduce the cupping artifacts as well. Moreover, it is able to reconstruct the anatomical structures outside the FOV. For example, the ribs on the left side are well reconstructed by the U-Net in Fig.\,\ref{Fig:TruncationNoiseFreeLW}(k). 
Nevertheless, the proposed DCR method further improves the accuracy of the U-Net reconstructions, achieving the smallest RMSE values of 49\,HU and 42\,HU for (f) and (l), respectively.

\begin{figure*}[h!]
\centering
\begin{minipage}{0.16\linewidth}
\centerline{$\boldsymbol{f}_{\text{reference}}$}
\smallskip
\end{minipage}
\begin{minipage}{0.16\linewidth}
\centerline{$\boldsymbol{f}_{\text{FBP}}$}
\smallskip
\end{minipage}
\begin{minipage}{0.16\linewidth}
\centerline{$\boldsymbol{f}_{\text{WCE}}$}
\smallskip
\end{minipage}
\begin{minipage}{0.16\linewidth}
\centerline{$\boldsymbol{f}_{\text{wTV}}$}
\smallskip
\end{minipage}
\begin{minipage}{0.16\linewidth}
\centerline{$\boldsymbol{f}_{\text{U-Net}}$}
\smallskip
\end{minipage}
\begin{minipage}{0.16\linewidth}
\centerline{$\boldsymbol{f}_{\text{DCR}}$}
\smallskip
\end{minipage}

\begin{minipage}{0.16\linewidth}
\subfigure[]
{
\includegraphics[width = \linewidth]{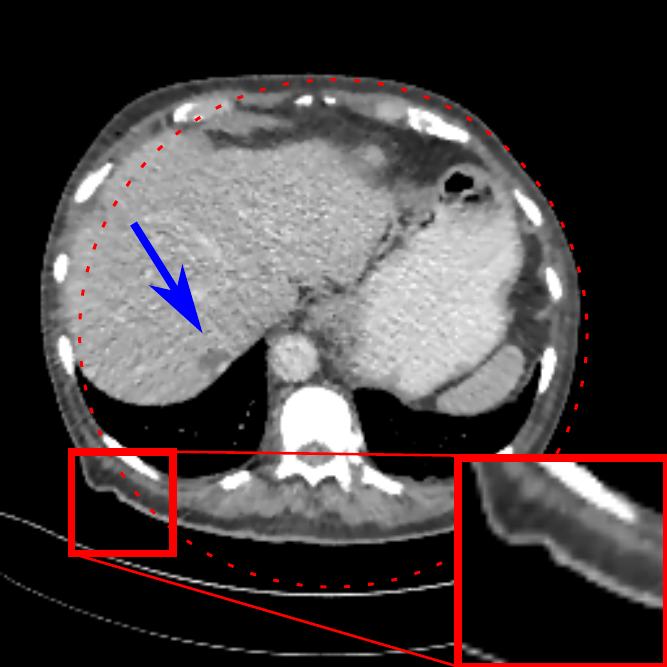}
}
\end{minipage}
\begin{minipage}{0.16\linewidth}
\subfigure[187\,HU]
{
\includegraphics[width = \linewidth]{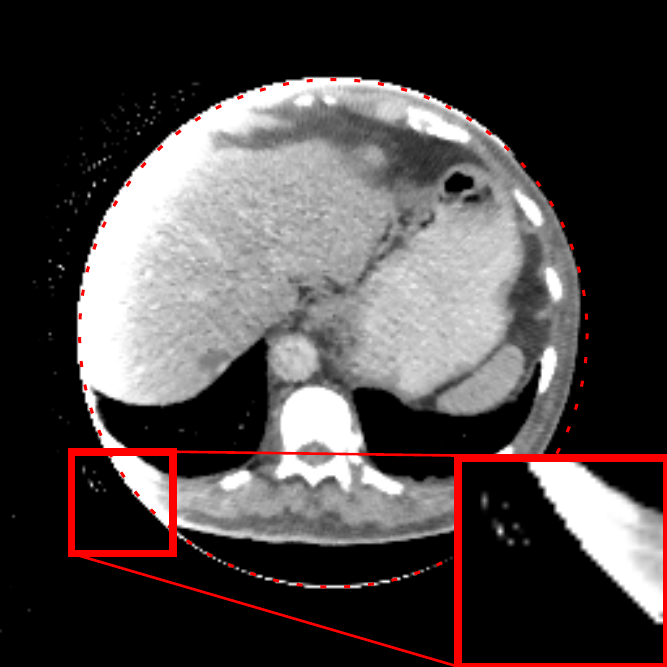}
}
\end{minipage}
\begin{minipage}{0.16\linewidth}
\subfigure[55\,HU]
{
\includegraphics[width = \linewidth]{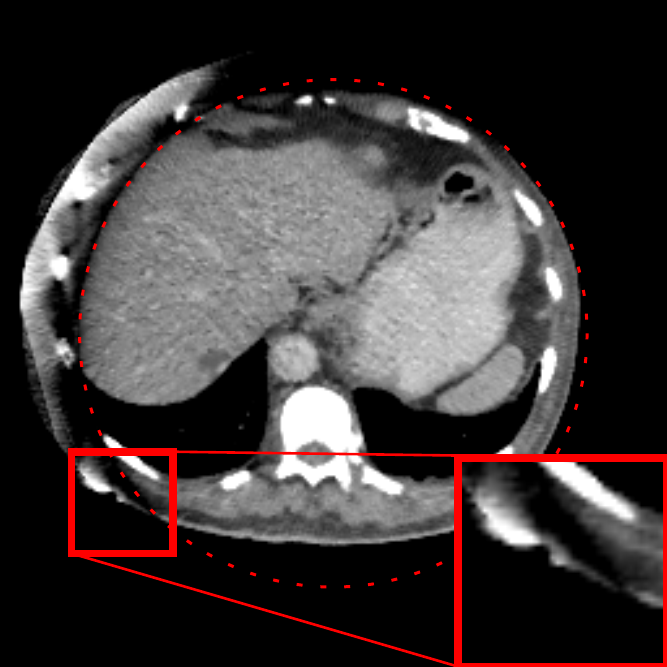}
}
\end{minipage}
\begin{minipage}{0.16\linewidth}
\subfigure[44\,HU]
{
\includegraphics[width = \linewidth]{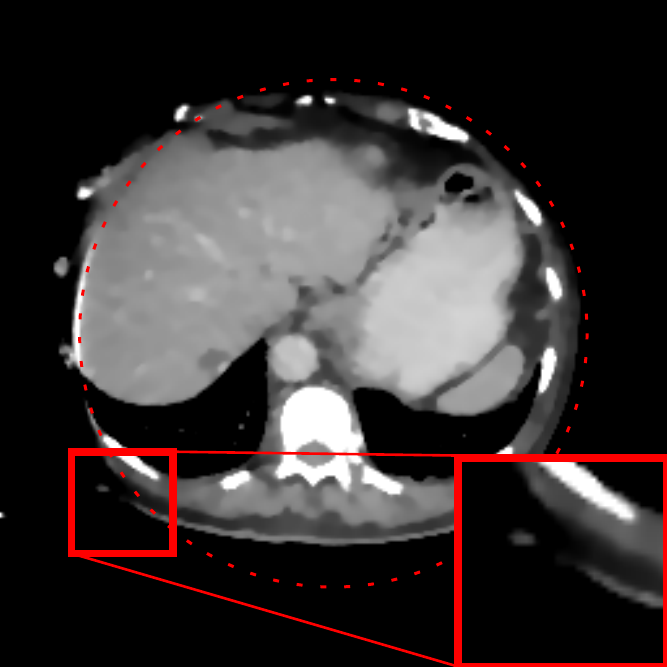}
}
\end{minipage}
\begin{minipage}{0.16\linewidth}
\subfigure[28\,HU]
{
\includegraphics[width = \linewidth]{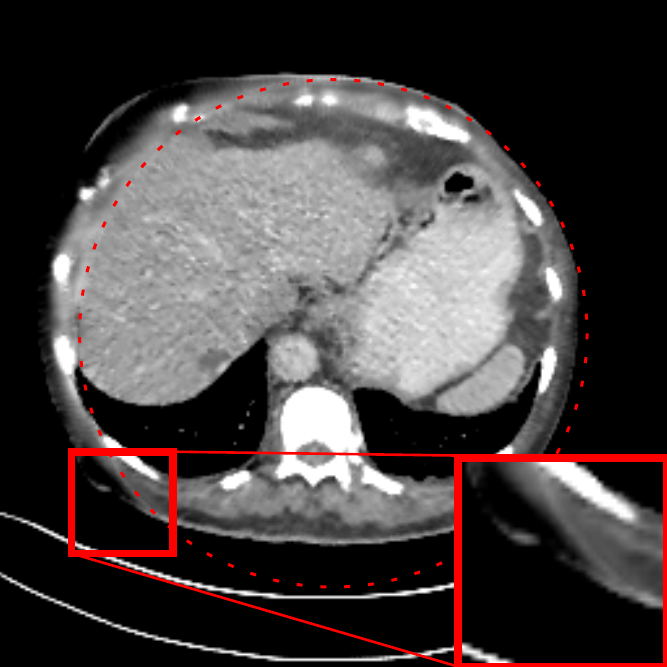}
}
\end{minipage}
\begin{minipage}{0.16\linewidth}
\subfigure[19\,HU]
{
\includegraphics[width = \linewidth]{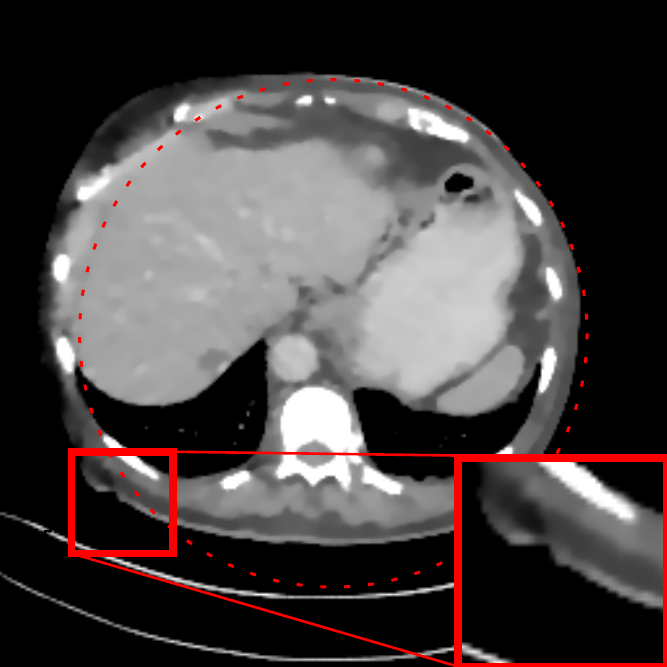}
}
\end{minipage}

\begin{minipage}{0.16\linewidth}
\subfigure[]
{
\includegraphics[width = \linewidth]{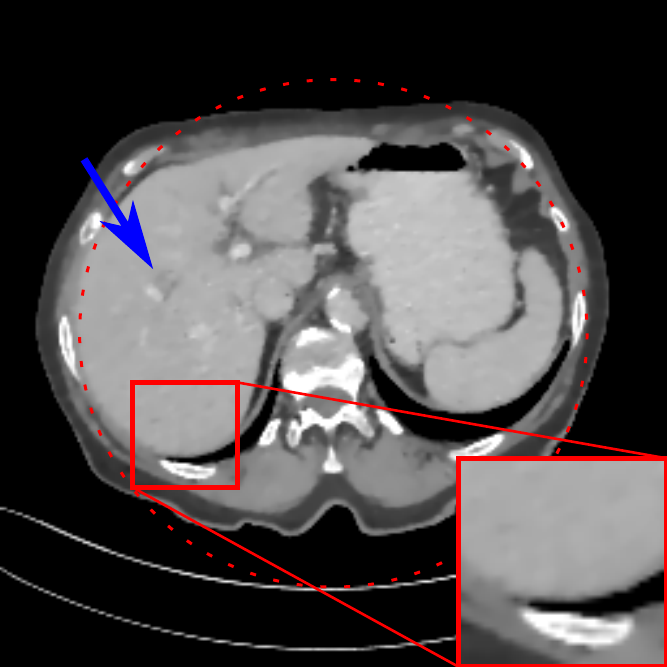}
}
\end{minipage}
\begin{minipage}{0.16\linewidth}
\subfigure[160\,HU]
{
\includegraphics[width = \linewidth]{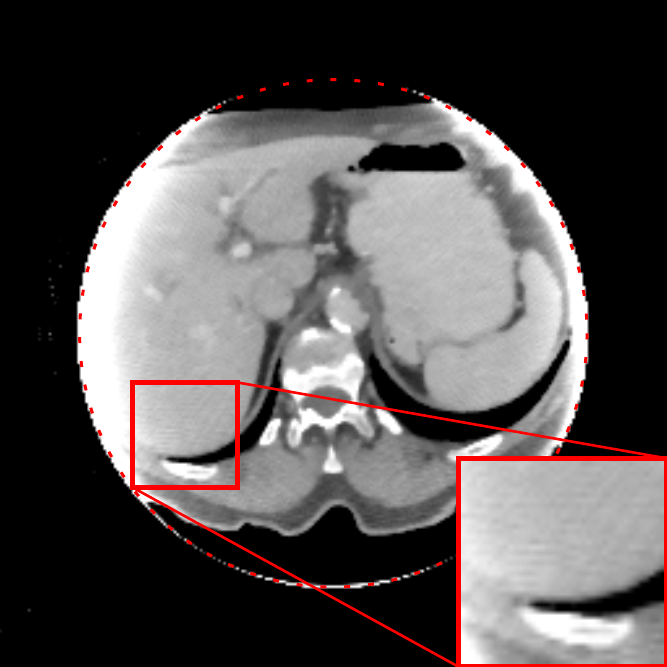}
}
\end{minipage}
\begin{minipage}{0.16\linewidth}
\subfigure[47\,HU]
{
\includegraphics[width = \linewidth]{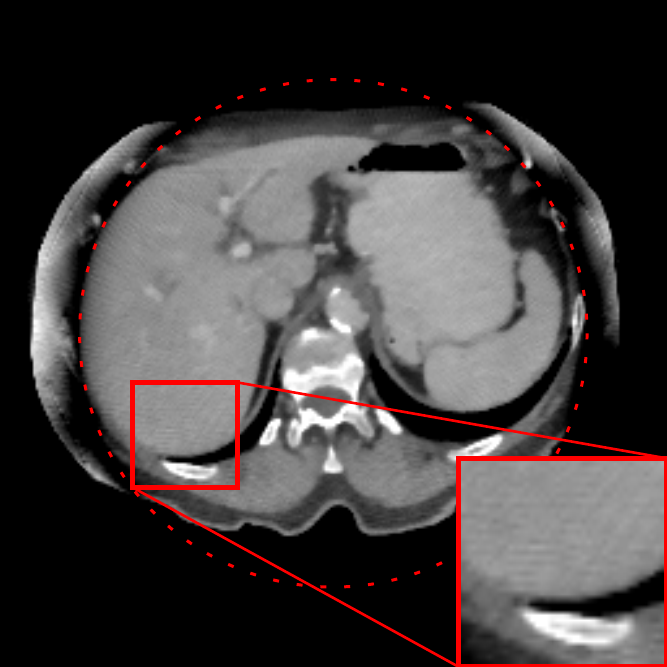}
}
\end{minipage}
\begin{minipage}{0.16\linewidth}
\subfigure[27\,HU]
{
\includegraphics[width = \linewidth]{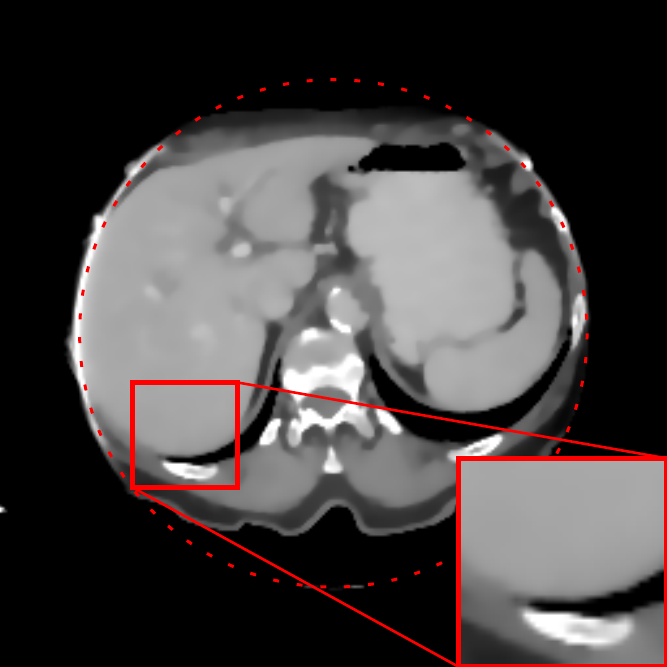}
}
\end{minipage}
\begin{minipage}{0.16\linewidth}
\subfigure[50\,HU]
{
\includegraphics[width = \linewidth]{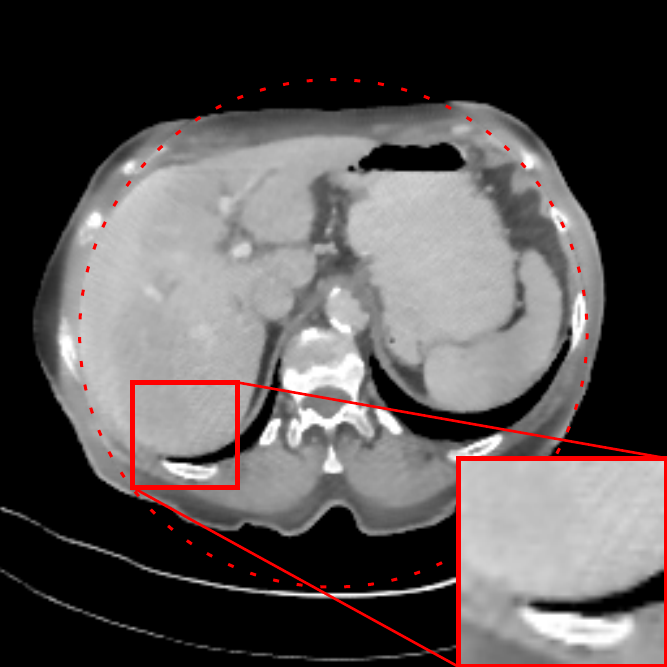}
}
\end{minipage}
\begin{minipage}{0.16\linewidth}
\subfigure[15\,HU]
{
\includegraphics[width = \linewidth]{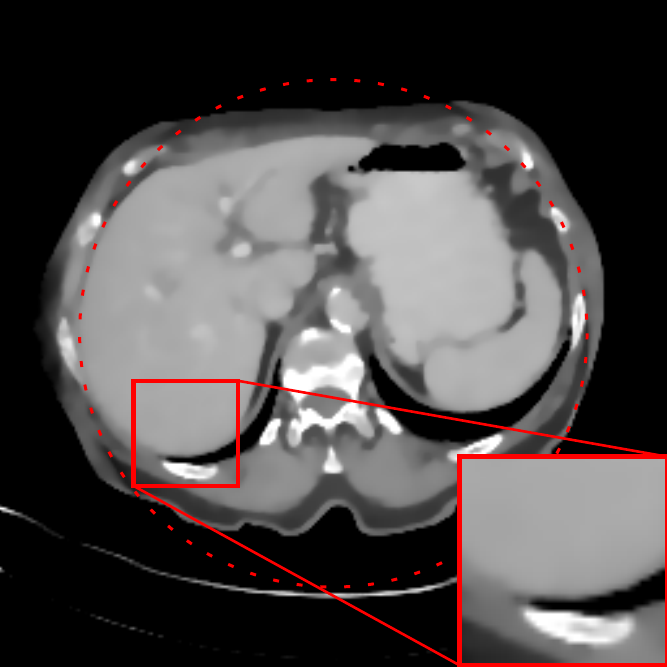}
}
\end{minipage}
\caption{Results of the two example slices in Fig.\,\ref{Fig:TruncationNoiseFreeLW} redisplayed in a narrow window of [-200, 200]\,HU. The FOV is indicated by the red dash circle. The lesions indicated by the blue arrows in (a) and (g) are of interest in the noisy case in Fig.\,\ref{Fig:truncationCorrectionNoisyCase}. The ROIs in (e) and (k) reconstructed by the U-Net have incorrect intensity values, which are improved by DCR in (f) and (l). The RMSE value inside the FOV for each method is displayed at the corresponding subcaption. }
\label{Fig:TruncationNoiseFreeSW}
\end{figure*}

The above two example slices are redisplayed in a narrow window of [-200, 200]\,HU in Fig.\,\ref{Fig:TruncationNoiseFreeSW}. The RMSE inside the FOV for each method is displayed in the corresponding subcaption. In this window, the detail structures of the liver can be observed, including the lesions indicated by the arrows in Fig.\,\ref{Fig:TruncationNoiseFreeSW}(a) and (g). In the FBP reconstructions displayed in Figs.\,\ref{Fig:TruncationNoiseFreeSW}(b) and (h), the cupping artifacts are clearly present, which are mitigated by WCE, wTV, U-Net and DCR, as the RMSE inside the FOV is significantly reduced. Among these methods, DCR achieves the lowest RMSE values of 19\,HU and 15\,HU respectively inside the FOV for the two slices. Although the U-Net reconstructs anatomical structures outside the FOV, these structures may not have accurate intensity values. For example, the tissue at the center of the ROI in Fig.\,\ref{Fig:TruncationNoiseFreeSW}(e) has low intensity. 
In contrast, it is observed better in the DCR reconstruction in Fig.\,\ref{Fig:TruncationNoiseFreeSW}(f). In addition, some structures inside the FOV may also have incorrect intensity values. For example, the ROI in Fig.\,\ref{Fig:TruncationNoiseFreeSW}(k) is brighter than that in the reference image (Fig.\,\ref{Fig:TruncationNoiseFreeSW}(g)), which is caused by the remaining cupping artifacts. This intensity bias is corrected in the DCR reconstruction in Fig.\,\ref{Fig:TruncationNoiseFreeSW}(l), as indicated in the ROI. 

\begin{figure*}[h]
\centering
\begin{minipage}{0.16\linewidth}
\centerline{$\boldsymbol{f}_{\text{reference}}$}
\smallskip
\end{minipage}
\begin{minipage}{0.16\linewidth}
\centerline{$\boldsymbol{f}_{\text{FBP}}$}
\smallskip
\end{minipage}
\begin{minipage}{0.16\linewidth}
\centerline{$\boldsymbol{f}_{\text{WCE}}$}
\smallskip
\end{minipage}
\begin{minipage}{0.16\linewidth}
\centerline{$\boldsymbol{f}_{\text{wTV}}$}
\smallskip
\end{minipage}
\begin{minipage}{0.16\linewidth}
\centerline{$\boldsymbol{f}_{\text{U-Net}}$}
\smallskip
\end{minipage}
\begin{minipage}{0.16\linewidth}
\centerline{$\boldsymbol{f}_{\text{DCR}}$}
\smallskip
\end{minipage}

\begin{minipage}{0.16\linewidth}
\subfigure[]
{
\includegraphics[width = \linewidth]{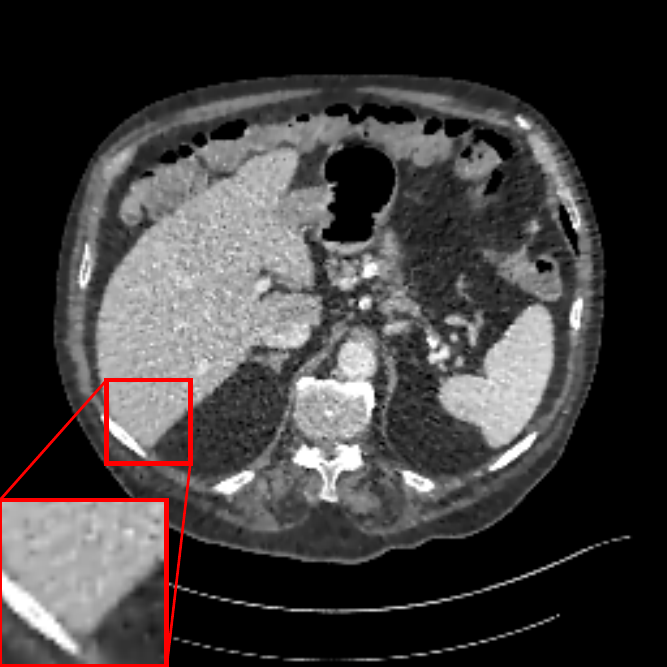}
}
\end{minipage}
\begin{minipage}{0.16\linewidth}
\subfigure[169\,HU]
{
\includegraphics[width = \linewidth]{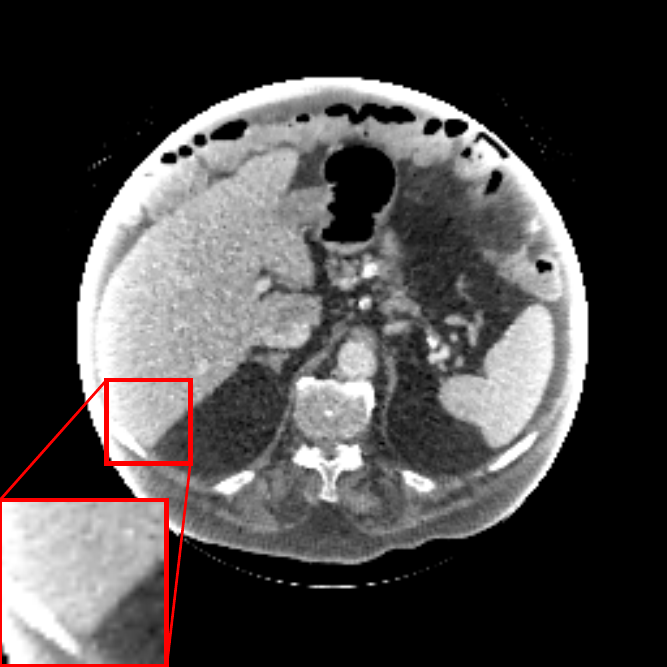}
}
\end{minipage}
\begin{minipage}{0.16\linewidth}
\subfigure[59\,HU]
{
\includegraphics[width = \linewidth]{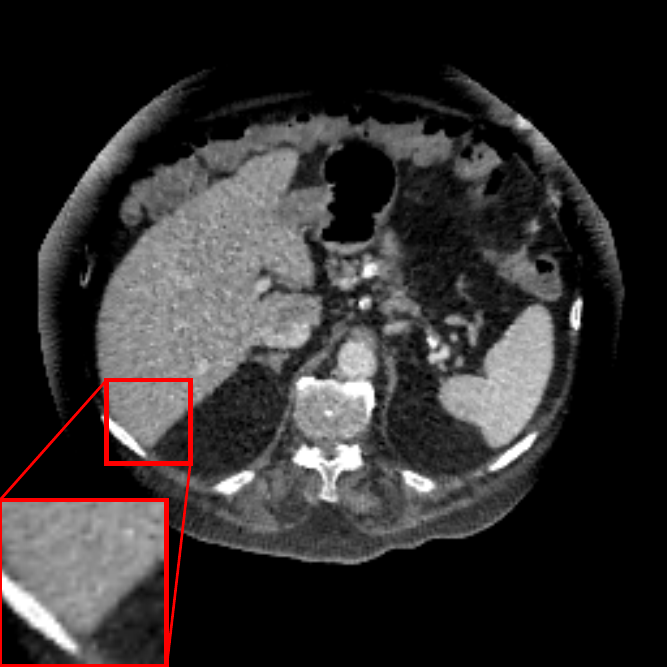}
}
\end{minipage}
\begin{minipage}{0.16\linewidth}
\subfigure[44\,HU]
{
\includegraphics[width = \linewidth]{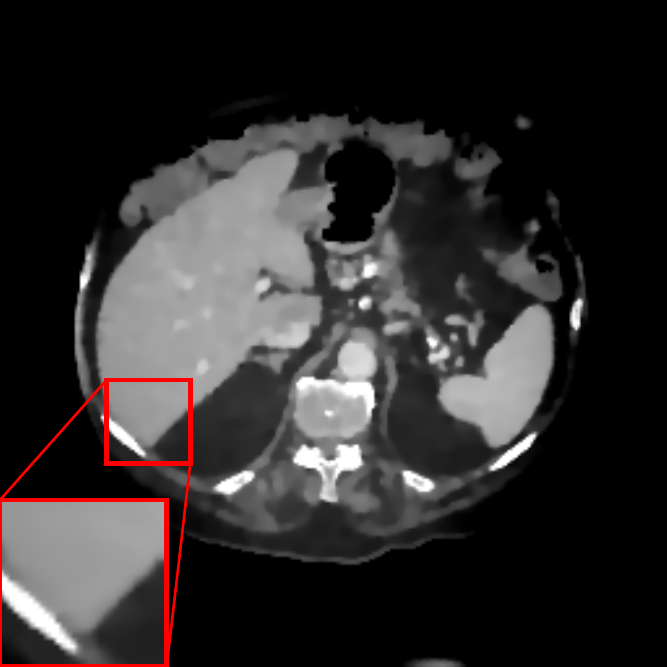}
}
\end{minipage}
\begin{minipage}{0.16\linewidth}
\subfigure[36\,HU]
{
\includegraphics[width = \linewidth]{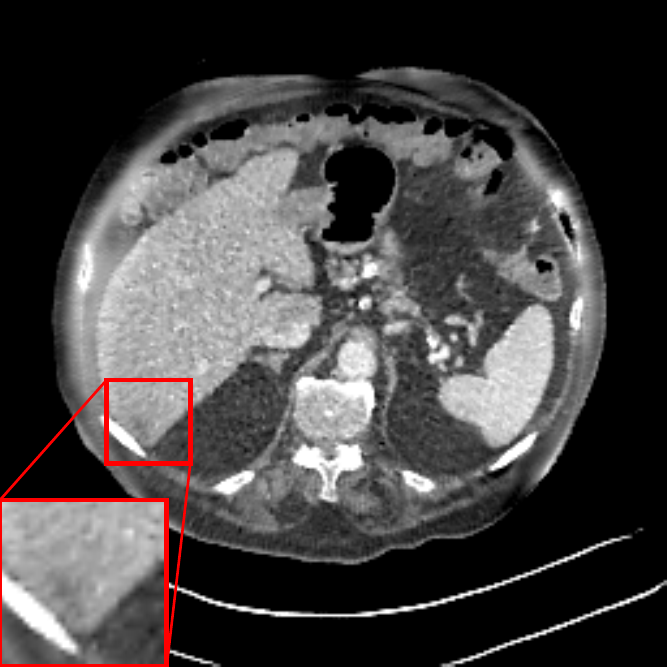}
}
\end{minipage}
\begin{minipage}{0.16\linewidth}
\subfigure[25\,HU]
{
\includegraphics[width = \linewidth]{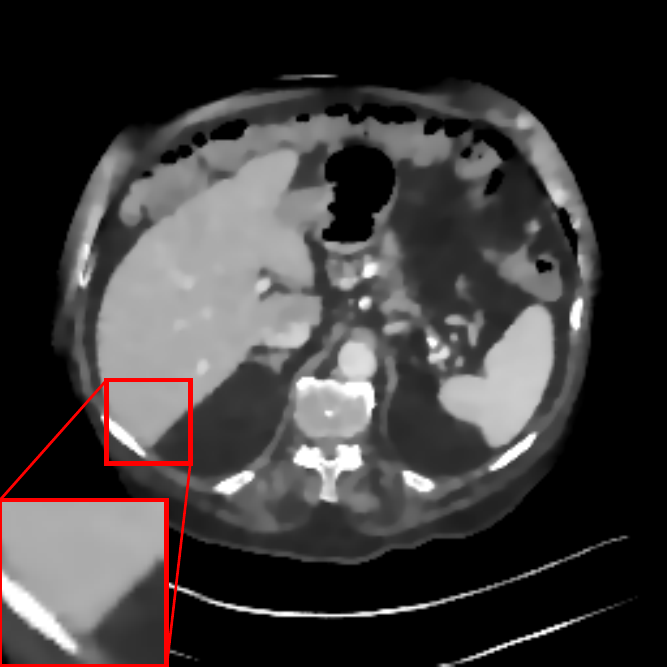}
}
\end{minipage}

\begin{minipage}{0.16\linewidth}
\subfigure[]
{
\includegraphics[width = \linewidth]{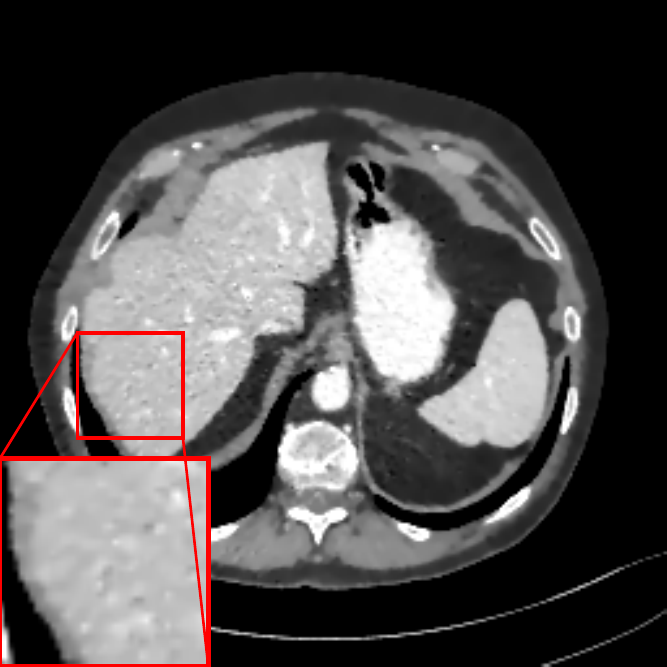}
}
\end{minipage}
\begin{minipage}{0.16\linewidth}
\subfigure[212\,HU]
{
\includegraphics[width = \linewidth]{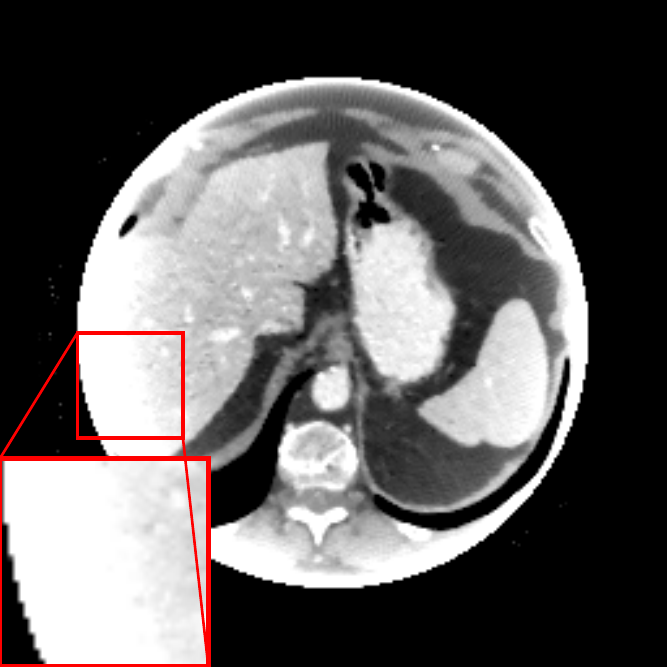}
}
\end{minipage}
\begin{minipage}{0.16\linewidth}
\subfigure[64\,HU]
{
\includegraphics[width = \linewidth]{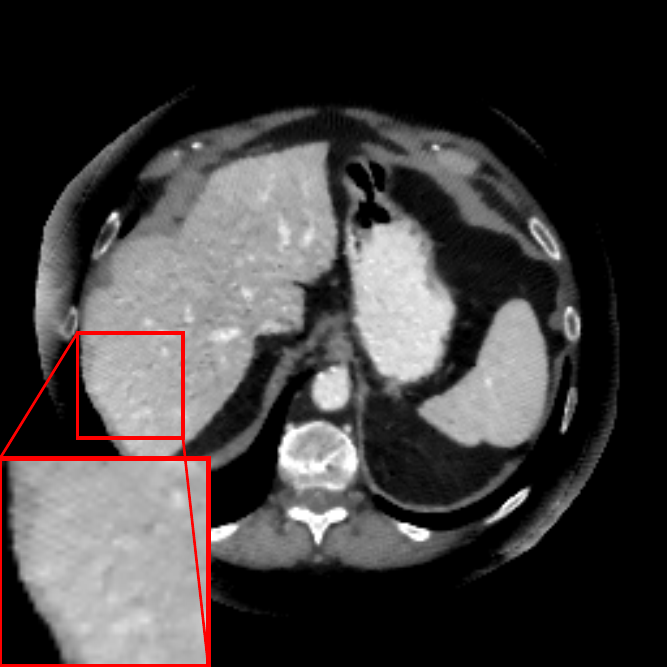}
}
\end{minipage}
\begin{minipage}{0.16\linewidth}
\subfigure[46\,HU]
{
\includegraphics[width = \linewidth]{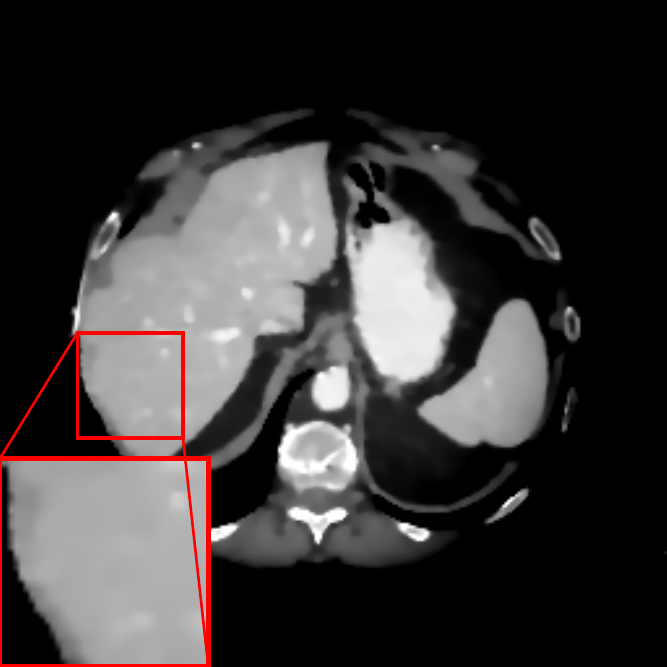}
}
\end{minipage}
\begin{minipage}{0.16\linewidth}
\subfigure[39\,HU]
{
\includegraphics[width = \linewidth]{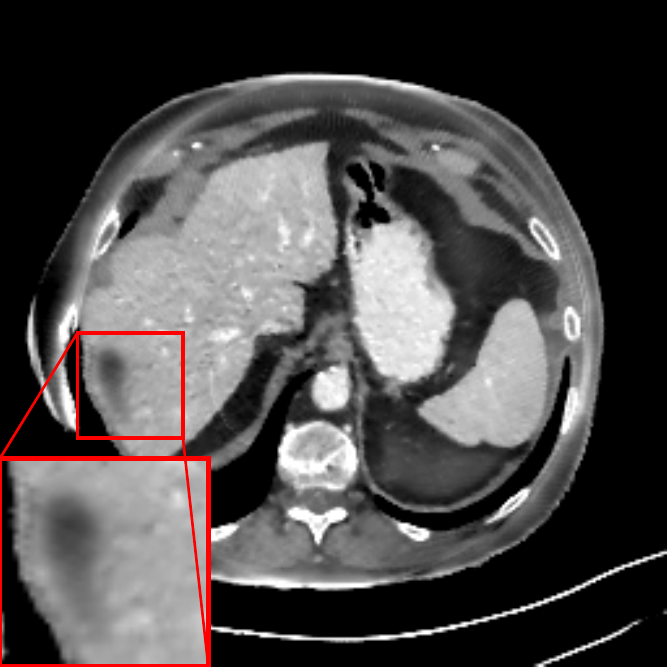}
}
\end{minipage}
\begin{minipage}{0.16\linewidth}
\subfigure[13\,HU]
{
\includegraphics[width = \linewidth]{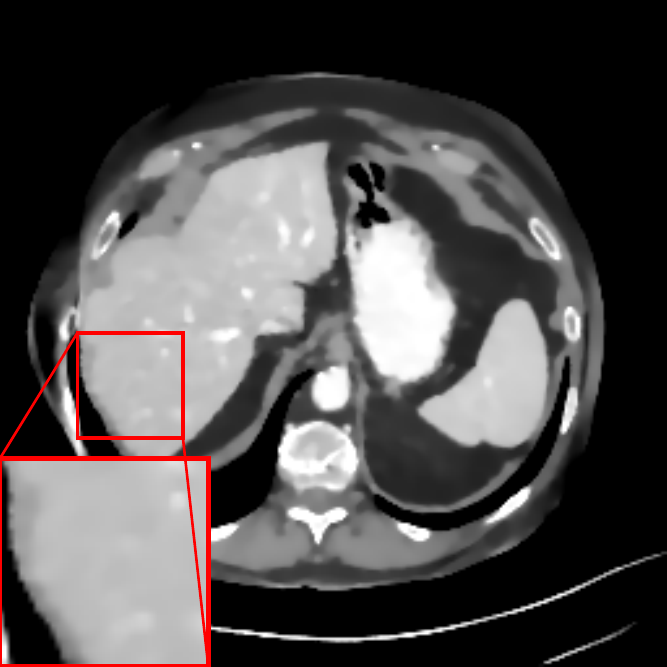}
}
\end{minipage}
\caption{False positive lesion examples reconstructed by the U-Net in the ROIs of (e) and (k) in the noise-free case, which are corrected by DCR in (f) and (l), window: [-200, 200]\,HU. The top and bottom rows are for the 187th slice of Patient NO.\,2 and the 123th slice of Patient NO.\,8, respectively. The RMSE value inside the FOV for each method is displayed at the corresponding subcaption.}
\label{Fig:FakeLesions}
\end{figure*}

The lesions in Fig.\,\ref{Fig:TruncationNoiseFreeSW} are reconstructed by all the methods. However, not all lesions observed in the U-Net reconstruction are reliable, even in the noise-free case. Two example results are displayed in Fig.\,\ref{Fig:FakeLesions}. In the ROI of Fig.\,\ref{Fig:FakeLesions}(e), a lesion is located at the bottom tip of the liver, which looks very realistic. In Fig.\,\ref{Fig:FakeLesions}(k), a large lesion is clearly visible. However, in the reference ROIs in Figs.\,\ref{Fig:FakeLesions}(a) and (g), these two lesions do not exist. With our proposed data consistency constraint, these fake lesions are removed, as demonstrated in the ROIs of Figs.\,\ref{Fig:FakeLesions}(f) and (l).

\begin{figure*}[h]
\centering
\begin{minipage}{0.16\linewidth}
\centerline{$\boldsymbol{f}_{\text{reference}}$}
\smallskip
\end{minipage}
\begin{minipage}{0.16\linewidth}
\centerline{$\boldsymbol{f}_{\text{FBP}}$}
\smallskip
\end{minipage}
\begin{minipage}{0.16\linewidth}
\centerline{$\boldsymbol{f}_{\text{WCE}}$}
\smallskip
\end{minipage}
\begin{minipage}{0.16\linewidth}
\centerline{$\boldsymbol{f}_{\text{wTV}}$}
\smallskip
\end{minipage}
\begin{minipage}{0.16\linewidth}
\centerline{$\boldsymbol{f}_{\text{U-Net}}$}
\smallskip
\end{minipage}
\begin{minipage}{0.16\linewidth}
\centerline{$\boldsymbol{f}_{\text{DCR}}$}
\smallskip
\end{minipage}

\begin{minipage}{0.16\linewidth}
\subfigure[]
{
\includegraphics[width = \linewidth]{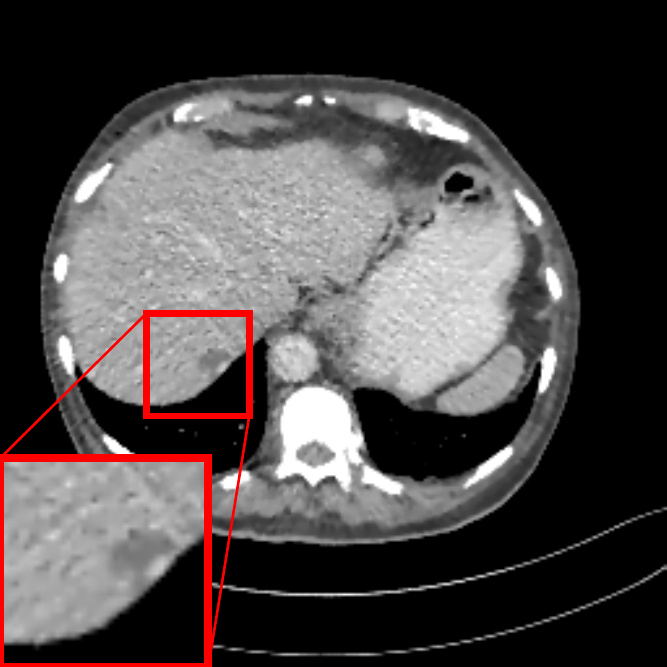}
}
\end{minipage}
\begin{minipage}{0.16\linewidth}
\subfigure[201\,HU]
{
\includegraphics[width = \linewidth]{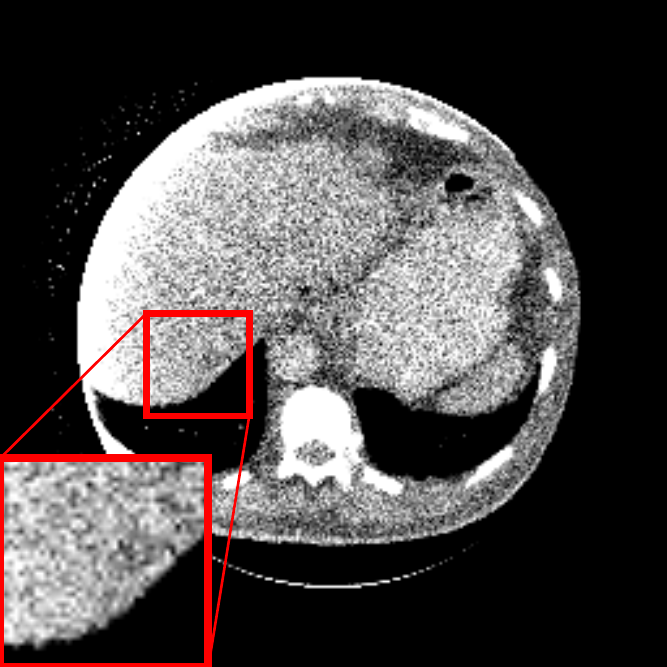}
}
\end{minipage}
\begin{minipage}{0.16\linewidth}
\subfigure[94\,HU]
{
\includegraphics[width = \linewidth]{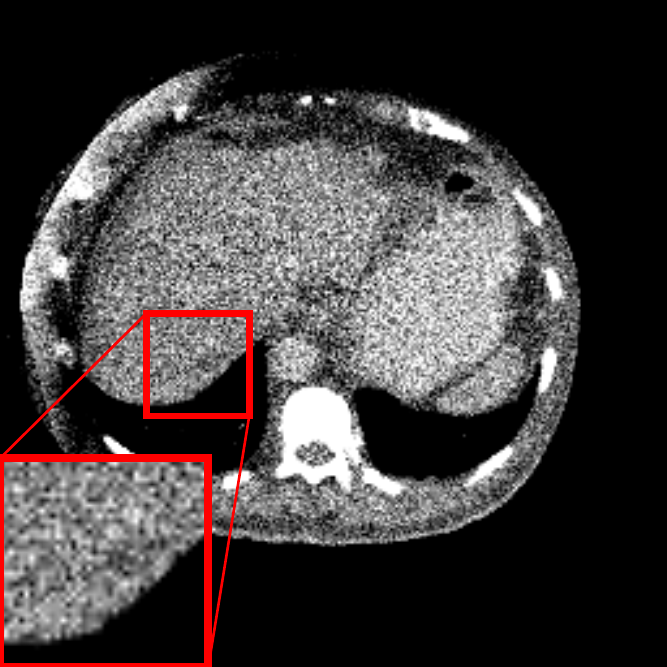}
}
\end{minipage}
\begin{minipage}{0.16\linewidth}
\subfigure[47\,HU]
{
\includegraphics[width = \linewidth]{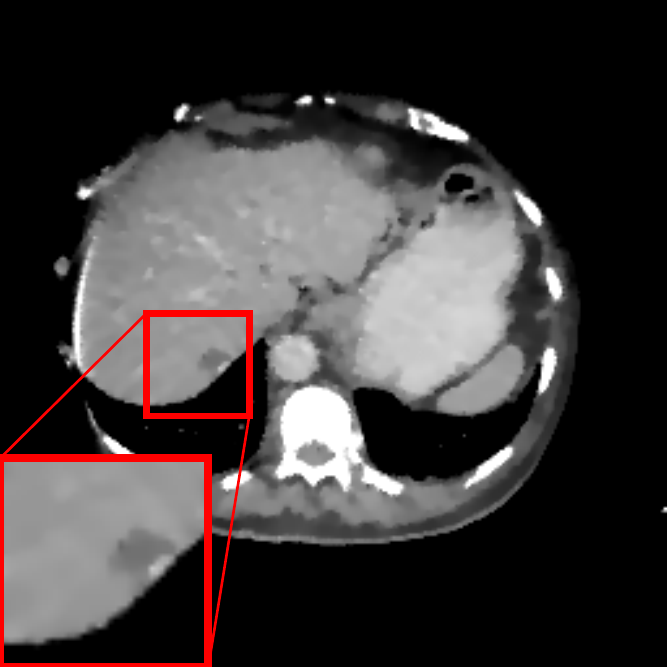}
}
\end{minipage}
\begin{minipage}{0.16\linewidth}
\subfigure[88\,HU]
{
\includegraphics[width = \linewidth]{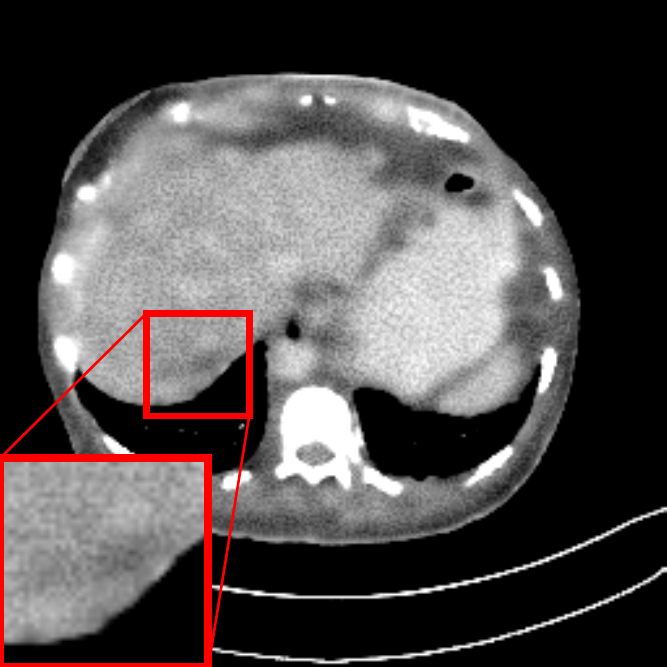}
}
\end{minipage}
\begin{minipage}{0.16\linewidth}
\subfigure[24\,HU]
{
\includegraphics[width = \linewidth]{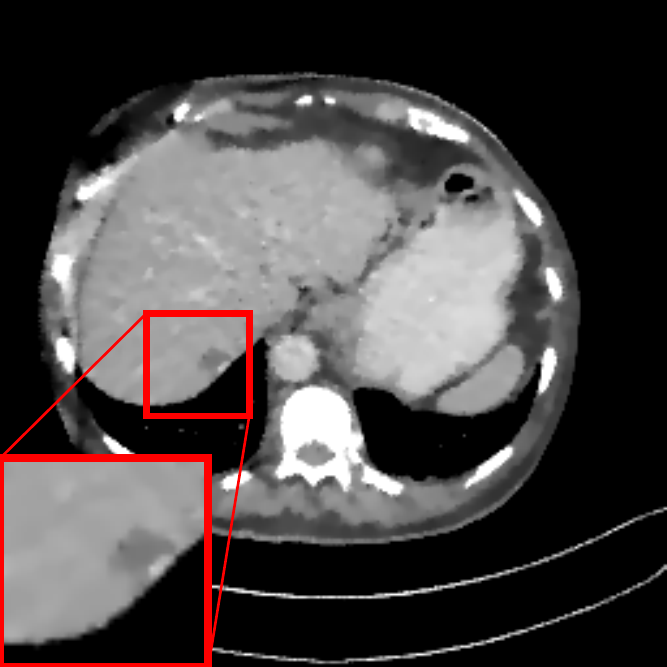}
}
\end{minipage}

\begin{minipage}{0.16\linewidth}
\subfigure[]
{
\includegraphics[width = \linewidth]{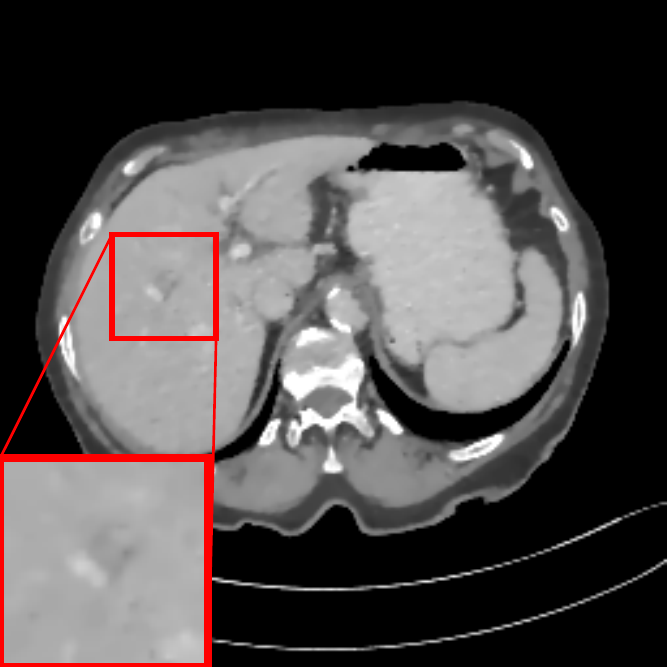}
}
\end{minipage}
\begin{minipage}{0.16\linewidth}
\subfigure[202\,HU]
{
\includegraphics[width = \linewidth]{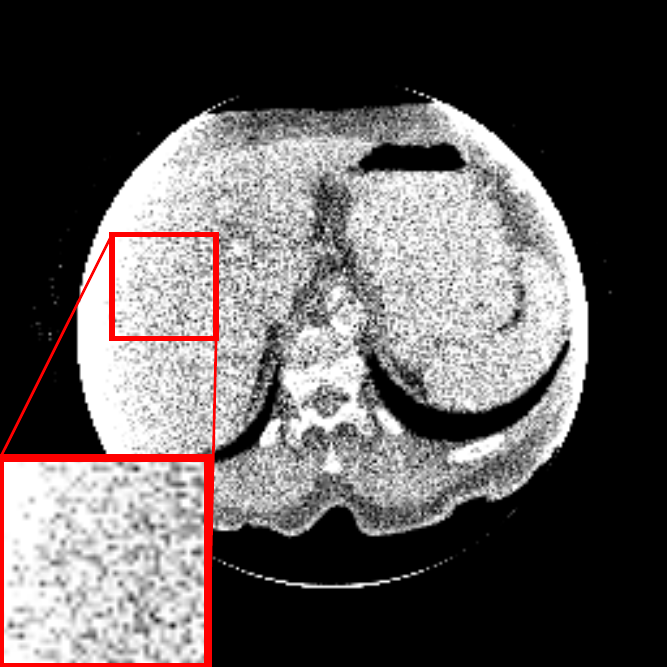}
}
\end{minipage}
\begin{minipage}{0.16\linewidth}
\subfigure[89\,HU]
{
\includegraphics[width = \linewidth]{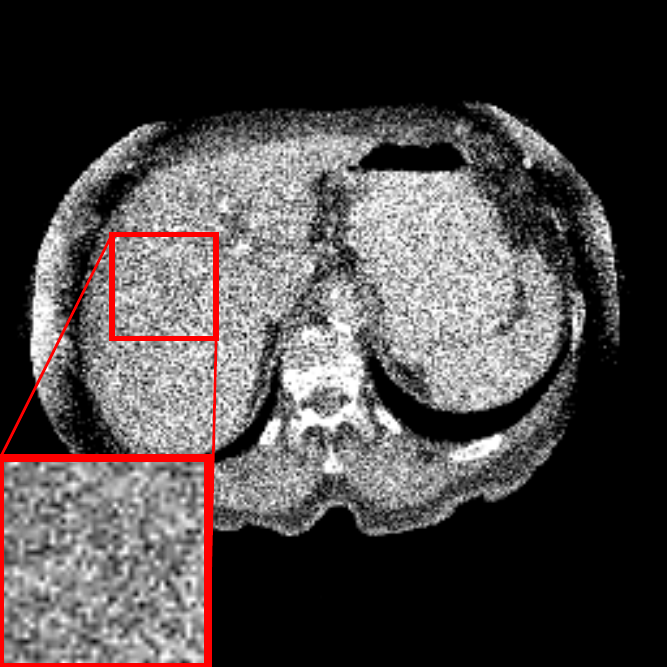}
}
\end{minipage}
\begin{minipage}{0.16\linewidth}
\subfigure[30\,HU]
{
\includegraphics[width = \linewidth]{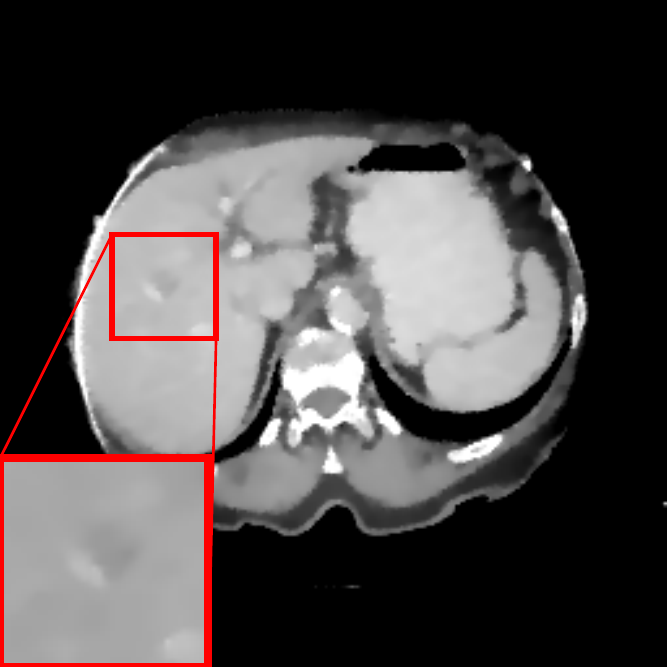}
}
\end{minipage}
\begin{minipage}{0.16\linewidth}
\subfigure[84\,HU]
{
\includegraphics[width = \linewidth]{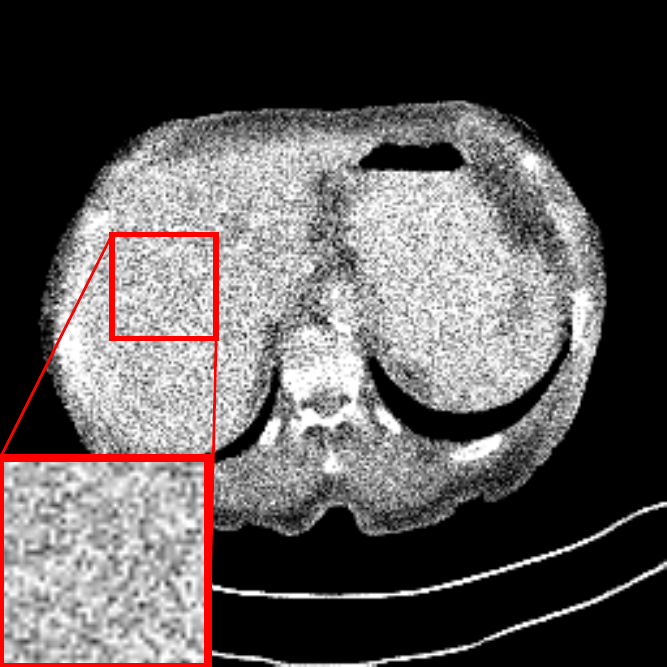}
}
\end{minipage}
\begin{minipage}{0.16\linewidth}
\subfigure[24\,HU]
{
\includegraphics[width = \linewidth]{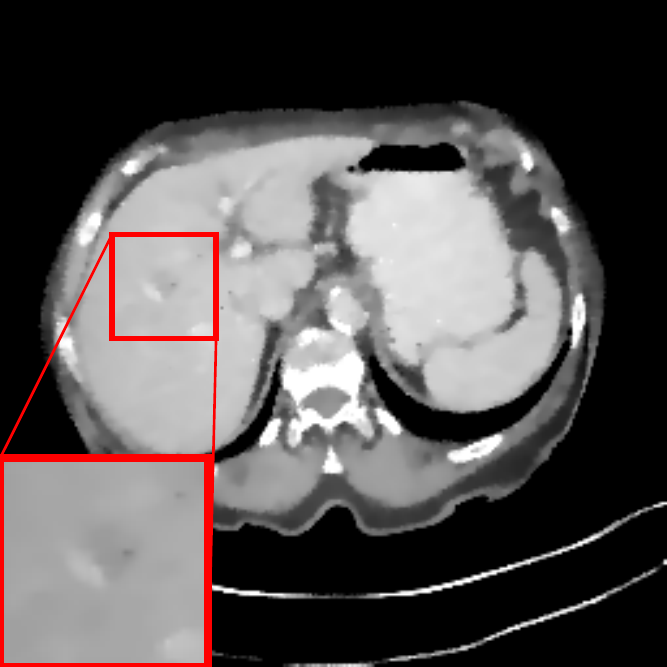}
}
\end{minipage}
\caption{Results of two example slices (the same two slices in Fig.\,\ref{Fig:TruncationNoiseFreeSW}) in the noisy case, window: [-200, 200]\,HU. The lesion in the ROI of (e) has incorrect outline and the very low contrast one in the ROI of (k) disappears entirely, while both of them are well reconstructed by DCR in (f) and (l). The RMSE value inside the FOV for each method is displayed at the corresponding subcaption.}
\label{Fig:truncationCorrectionNoisyCase}
\end{figure*}

The truncation correction methods are also evaluated in the noisy case. The results of two example slices (the same two slices in Fig.\,\ref{Fig:TruncationNoiseFreeSW}) are displayed in Fig.\,\ref{Fig:truncationCorrectionNoisyCase}. Noise patterns are observed in the FBP reconstruction (Figs.\,\ref{Fig:truncationCorrectionNoisyCase}(b) and (h)) and the WCE reconstruction (Figs.\,\ref{Fig:truncationCorrectionNoisyCase}(c) and (i)), where the lesions in the corresponding ROIs are obscured by noise.
Figs.\,\ref{Fig:truncationCorrectionNoisyCase}(d) and (j) demonstrate that wTV is able to reduce Poisson noise very well and reveal the lesion areas. 
For the U-Net result displayed in Fig.\,\ref{Fig:truncationCorrectionNoisyCase}(e), the lesion boundary in the ROI is very blurry and its area is much larger than that in the reference ROI of Fig.\,\ref{Fig:truncationCorrectionNoisyCase}(a). For the U-Net result in Fig.\,\ref{Fig:truncationCorrectionNoisyCase}(k), Poisson noise remains and the low contrast (about 20\,HU) lesion in the ROI is obscured by the severe noise. These observations indicate that deep learning is not robust enough in the existence of noise, even if the U-Net is trained on data with Poisson noise. The proposed DCR method combines the advantages of wTV and U-Net. It reduces both the cupping artifacts and the Poisson noise inside the FOV, preserving organ details. Meanwhile, it reconstructs the anatomical structures outside the FOV, as demonstrated in Figs.\,\ref{Fig:truncationCorrectionNoisyCase}(f) and (l). Especially, both lesions in Figs.\,\ref{Fig:truncationCorrectionNoisyCase}(f) and (l) are reconstructed by DCR. Among all the algorithms, it achieves the lowest RMSE value of 24\,HU inside the FOV.

\begin{table*}[h]
\caption{The quantitative evaluation results of different methods for truncation correction in the noisy case.}
\label{tab:FOVNoisy}
\centering
\begin{tabular}{|l|ccc|ccc|ccc|ccc|}
\hline
\ & \multicolumn{3}{|c|}{RMSE inside FOV} & \multicolumn{3}{|c|}{RMSE} & \multicolumn{3}{|c|}{SSIM inside FOV}& \multicolumn{3}{|c|}{SSIM}\\
\hline
Patient & wTV & U-Net & DCR & wTV & U-Net & DCR & wTV & U-Net & DCR & wTV & U-Net & DCR\\
\hline
NO.\,1 & 27\,HU & 46\,HU & \textbf{23\,HU} & 105\,HU & 80\,HU & \textbf{54\,HU} & 0.999 &0.996 &0.999 &0.975 & 0.986 & \textbf{0.993}\\
\hline
NO.\,2 & 66\,HU & 54\,HU & \textbf{27\,HU} & 202\,HU & 106\,HU & \textbf{80\,HU} &0.993 &0.995 &0.999 & 0.889 & 0.975 & \textbf{0.985}\\
\hline
NO.\,3 & 51\,HU & 61\,HU & \textbf{28\,HU} & 163\,HU & 112\,HU & \textbf{80\,HU} &0.996 &0.994 &0.999 & 0.940 & 0.975 & \textbf{0.986}\\
\hline
NO.\,4 & 26\,HU & 48\,HU & \textbf{22\,HU} & 114\,HU & 91\,HU & \textbf{72\,HU} &0.999 &0.997 & \textbf{0.999} & 0.976 & 0.984 & \textbf{0.990}\\
\hline
NO.\,5 & 48\,HU & 57\,HU & \textbf{29\,HU} & 193\,HU & 143\,HU & \textbf{116\,HU} &0.996 &0.994 &\textbf{0.999} & 0.900 & 0.955 & \textbf{0.969}\\
\hline
NO.\,6 & 30\,HU & 50\,HU & \textbf{28\,HU} & 150\,HU & 135\,HU & \textbf{107\,HU} &0.998 &0.995 &\textbf{0.998} & 0.946 & 0.957 & \textbf{0.973}\\
\hline
NO.\,7 & 26\,HU & 54\,HU & \textbf{22\,HU} & 129\,HU & 125\,HU & \textbf{89\,HU} &0.999 &0.995 &\textbf{0.999} & 0.963 & 0.967 & \textbf{0.982}\\
\hline
NO.\,8 & 76\,HU & 50\,HU & \textbf{25\,HU} & 213\,HU & 122\,HU & \textbf{93\,HU} &0.990 &0.996 &\textbf{0.999}& 0.875 & 0.968 & \textbf{0.980}\\
\hline
NO.\,9 & 23\,HU & 57\,HU & \textbf{21\,HU} & 100\,HU & 88\,HU & \textbf{64\,HU} &0.999 &0.994 &\textbf{0.999} & 0.977 & 0.983 & \textbf{0.991}\\
\hline
NO.\,10 & 46\,HU & 63\,HU & \textbf{22\,HU} & 139\,HU & 101\,HU & \textbf{74\,HU} &0.996 &0.993 &\textbf{0.999} & 0.950 & 0.976 & \textbf{0.986}\\
\hline
NO.\,11 & 20\,HU & 50\,HU & \textbf{18\,HU} & 96\,HU & 111\,HU & \textbf{82\,HU} &0.999 &0.996 &\textbf{0.999}& 0.982 & 0.972 & \textbf{0.983}\\
\hline
NO.\,12 & 24\,HU & 46\,HU & \textbf{18\,HU} & 59\,HU & 51\,HU & \textbf{27\,HU} &0.999 &0.996 &\textbf{0.999} & 0.993 & 0.995 & \textbf{0.999}\\
\hline
NO.\,13 & 50\,HU & 61\,HU & \textbf{30\,HU} & 175\,HU & 144\,HU & \textbf{127\,HU} &0.995 &0.992 &\textbf{0.998} & 0.922 & 0.952 & \textbf{0.962}\\
\hline
NO.\,14 & 60\,HU & 60\,HU & \textbf{24\,HU} & 179\,HU & 114\,HU & \textbf{88\,HU} &0.994 &0.994 &\textbf{0.999} & 0.917 & 0.971 & \textbf{0.982}\\
\hline
NO.\,15 & 20\,HU & 66\,HU & \textbf{18\,HU} & 77\,HU & 94\,HU & \textbf{58\,HU} &0.999 &0.993 &\textbf{0.999}& 0.985 & 0.977 & \textbf{0.991}\\
\hline
NO.\,16 & 27\,HU & 45\,HU & \textbf{17\,HU} & 84\,HU & 79\,HU & \textbf{61\,HU} &0.999 &0.997 &\textbf{1.000} &0.986 & 0.988 & \textbf{0.993}\\
\hline
NO.\,17 & 29\,HU & 75\,HU & \textbf{24\,HU} & 116\,HU & 99\,HU & \textbf{66\,HU} &0.999 &0.991 &\textbf{0.999}& 0.968 & 0.975 & \textbf{0.993}\\
\hline
Mean & 38\,HU & 55\,HU & \textbf{23\,HU} & 135\,HU & 106\,HU & \textbf{78\,HU} &0.997 &0.995 &\textbf{0.999}& 0.950 & 0.974 & \textbf{0.985}\\
\hline
Standard deviation & 17\,HU & 8\,HU & \textbf{4\,HU} & 47\,HU & 26\,HU & \textbf{24\,HU} &0.0027 &0.0017 &\textbf{0.0004}& 0.0368 & 0.0117 & \textbf{0.0095}\\
\hline
\end{tabular}
\end{table*}

The quantitative evaluation results of wTV, U-Net and DCR for different patients in the noisy case are displayed in Tab.\,\ref{tab:FOVNoisy}, where RMSE and SSIM are computed using either the FOV area or the whole image. For each patient, DCR achieves the best performance among the three methods. Particularly, for image quality inside the FOV, DCR achieves the lowest mean RMSE value of 23\,HU and the highest mean SSIM index of 0.999 for all the patients, while the U-Net achieves 55\,HU and 0.995 respectively for these two metrics. For the whole image quality, DCR achieves a low mean RMSE value of 78\,HU and a high mean SSIM index of 0.985.
\begin{figure*}
\centering

\begin{minipage}{0.16\linewidth}
\subfigure[$e_1=0.05, e_2=0.5$]
{
\includegraphics[width = \linewidth]{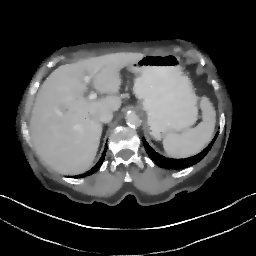}
}
\end{minipage}
\begin{minipage}{0.16\linewidth}
\subfigure[SART]
{
\includegraphics[width = \linewidth]{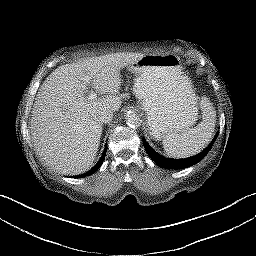}
}
\end{minipage}
\begin{minipage}{0.16\linewidth}
\subfigure[$e_1=0.005, e_2=0.5$]
{
\includegraphics[width = \linewidth]{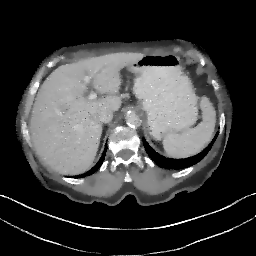}
}
\end{minipage}
\begin{minipage}{0.16\linewidth}
\subfigure[$e_1=0.5, e_2=0.5$]
{
\includegraphics[width = \linewidth]{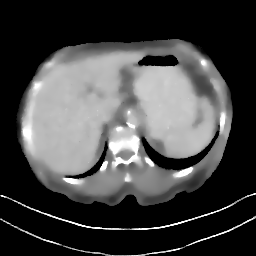}
}
\end{minipage}
\begin{minipage}{0.16\linewidth}
\subfigure[$e_1=0.05, e_2=0.1$]
{
\includegraphics[width = \linewidth]{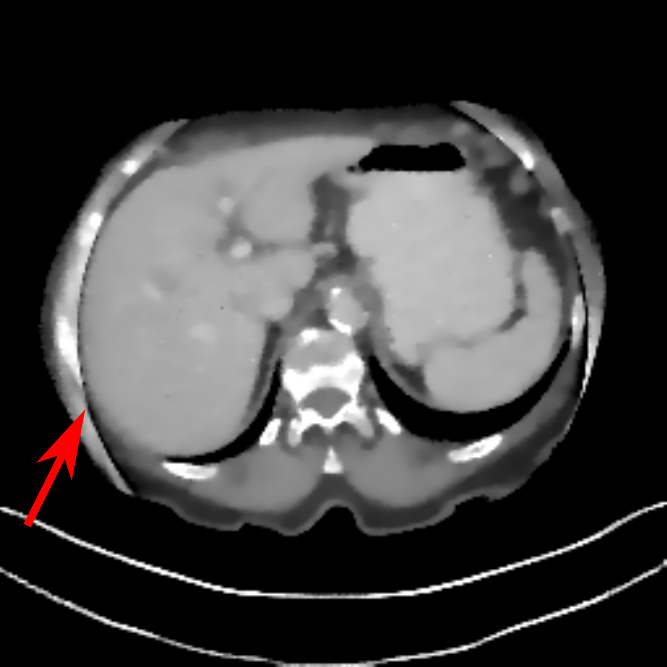}
}
\end{minipage}
\begin{minipage}{0.16\linewidth}
\subfigure[$e_1=0.05, e_2=5$]
{
\includegraphics[width = \linewidth]{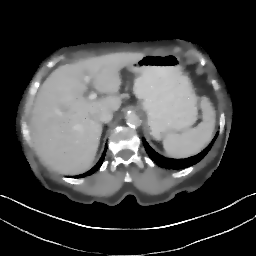}
}
\end{minipage}
\caption{Reconstruction results of the 150th slice of Patient NO.\,17 with different parameters, window: [-200, 200]\,HU. (a) $e_1=0.05, e_2=0.5$; (b) $e_1=0.05, e_2=0.5$ without wTV regularization, where Poisson noise remains; (c) $e_1=0.005, e_2=0.5$, where Poisson noise partially remains; (d) $e_1=0.5, e_2=0.5$, where fine structures are over smoothed; (e) $e_1=0.05, e_2=0.1$, where the FOV boundary area is incorrect; (f) $e_1=0.05, e_2=5$, no significant visual difference to (a). }
\label{Fig:paraSelection}
\end{figure*}

\begin{figure}
\centering
\includegraphics[width = 1\linewidth]{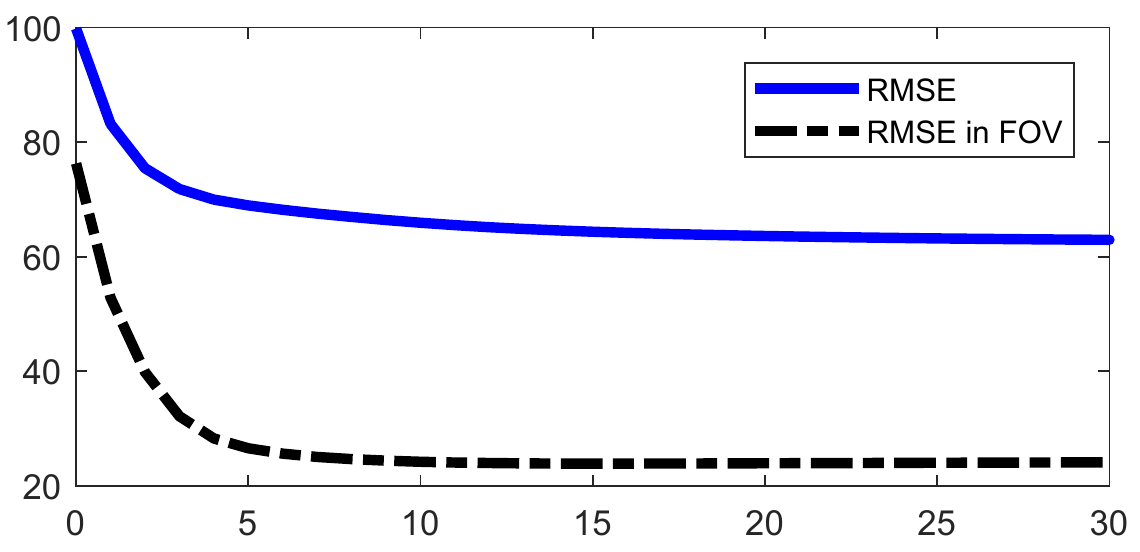}
\caption{The RMSE values for the whole body area and the area inside the FOV of Patient NO.\,17 over iterations.}
\label{Fig:iterationNumber}
\end{figure}

In the proposed DCR algorithm, $e_1$ and $e_2$ in Eqn.\,(\ref{eqn:ObjectiveDataConsistentDeepLearning}) are two important performance parameters. In the noise-free case, a small value close to zero is fine for $e_1$. However, in the noisy case, a proper value for $e_1$ needs to be set for noise tolerance. Therefore, several example images in the noisy case are displayed in Fig.\,\ref{Fig:paraSelection} to indicate the influence of these two parameters. All the images are obtained with other parameters of $\epsilon = 10$\,HU for weight update, $\lambda = 0.8$ for SART update and $n_{\max} = 10$ for the total number of iterations. Fig.\,\ref{Fig:paraSelection}(a) (the same as Fig.\,\ref{Fig:truncationCorrectionNoisyCase}(l)) is the result with the empirically chosen parameters $e_1 = 0.05$ and $e_2 = 0.5$. Fig.\,\ref{Fig:paraSelection}(b) is the result with the parameters $e_1 = 0.05$ and $e_2 = 0.5$ as well, but only with the SART update for data fidelity without the wTV minimization. As expected, Poisson noise remains in Fig.\,\ref{Fig:paraSelection}(b), indicating the necessity of the wTV minimization. Fig.\,\ref{Fig:paraSelection}(c) and (d) are the results with $e_1 = 0.005$ and $e_1 = 0.5$, respectively. In Fig.\,\ref{Fig:paraSelection}(c), a large portion of Poisson noise is reduced. However, still some Poisson noise remains due to the small value of $e_1$. In Fig.\,\ref{Fig:paraSelection}(d), the Poisson noise is entirely removed. However, fine structures are over-smoothed. These observations indicate that $e_1$ a parameter controlling the trade-off between noise reduction and high spatial resolution. Hence, $e_1 = 0.05$ is recommended empirically for the noisy case in this work. Regarding the parameter $e_2$, Fig.\,\ref{Fig:paraSelection}(e) and (f) are the results with $e_2 = 0.1$ and $e_2 = 5$, respectively. In Fig.\,\ref{Fig:paraSelection}(e), the FOV boundary is present, caused by the discontinuity between $\hat{\boldsymbol{p}}_{\text{u}}$ and $\boldsymbol{p}_{\text{m}}$ at the transition area. Fig.\,\ref{Fig:paraSelection}(f) is very similar to Fig.\,\ref{Fig:paraSelection}(a) where the pixel values at the FOV boundary are corrected while fine structures are preserved. This indicates that the selection of $e_2$ is relatively flexible as long as it is large enough. With the chosen parameters $e_1 = 0.05$ and $e_2 = 0.5$, the average RMSE values for the whole image and the area inside the FOV of Patient NO.\,17 are plotted over iterations in Fig.\,\ref{Fig:iterationNumber}. The RMSE inside the FOV converges faster that that of the whole image. However, both of them have little change after 10 iterations. Therefore, in this work, choosing the total iteration number $n_{\max} = 10$ is sufficient.

\subsection{Limited-Angle Tomography}

\begin{figure*}[tbh]
\centering
\begin{minipage}{0.16\linewidth}
\centering
$\vf_{\text{reference}}$
\end{minipage}
\begin{minipage}{0.16\linewidth}
\centering
$\vf_{\text{FBP}}$
\end{minipage}
\begin{minipage}{0.16\linewidth}
\centering
$\vf_{\text{wTV}}$
\end{minipage}
\begin{minipage}{0.16\linewidth}
\centering
$\vf_{\text{U-Net}}$
\end{minipage}
\begin{minipage}{0.16\linewidth}
\centering
$\vf_{\text{DCR}}$
\end{minipage}

\vspace{3pt}

\begin{minipage}{0.16\linewidth}
\subfigure[]{
\includegraphics[width = \linewidth]{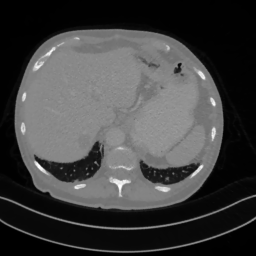}
}
\end{minipage}
\begin{minipage}{0.16\linewidth}
\subfigure[ 0.862]{
\includegraphics[width = \linewidth]{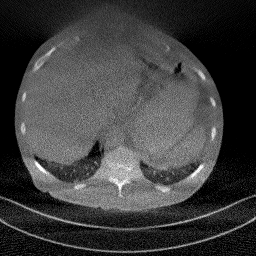}
}
\end{minipage}
\begin{minipage}{0.16\linewidth}
\subfigure[ 0.988]{
\includegraphics[width = \linewidth]{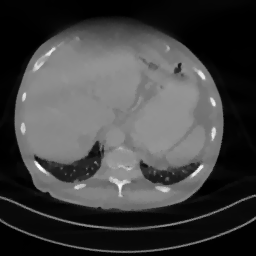}
}
\end{minipage}
\begin{minipage}{0.16\linewidth}
\subfigure[ 0.992]{
\includegraphics[width = \linewidth]{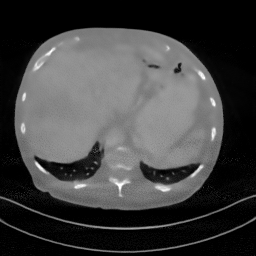}
}
\end{minipage}
\begin{minipage}{0.16\linewidth}
\subfigure[ 0.996]{
\includegraphics[width = \linewidth]{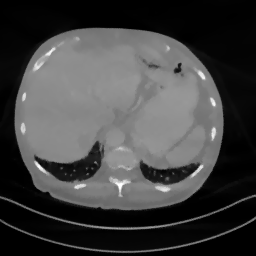}
}
\end{minipage}

\begin{minipage}{0.16\linewidth}
\subfigure[]{
\includegraphics[width = \linewidth]{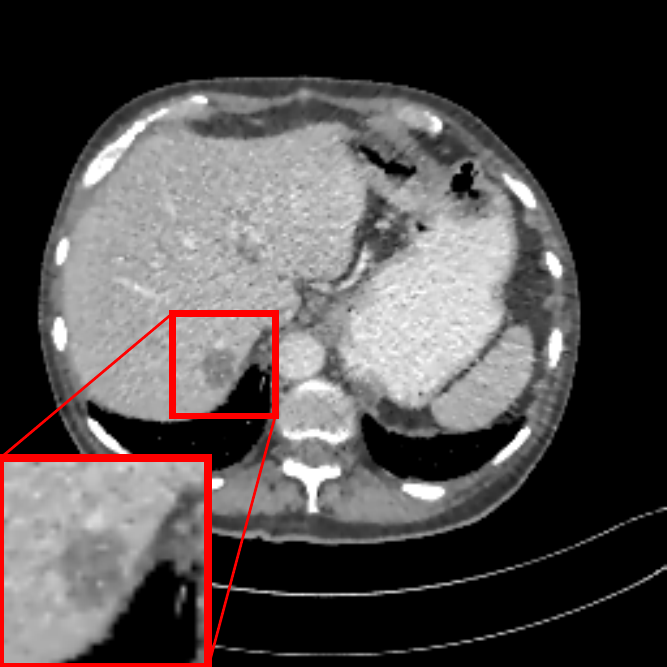}
}
\end{minipage}
\begin{minipage}{0.16\linewidth}
\subfigure[266\,HU]{
\includegraphics[width = \linewidth]{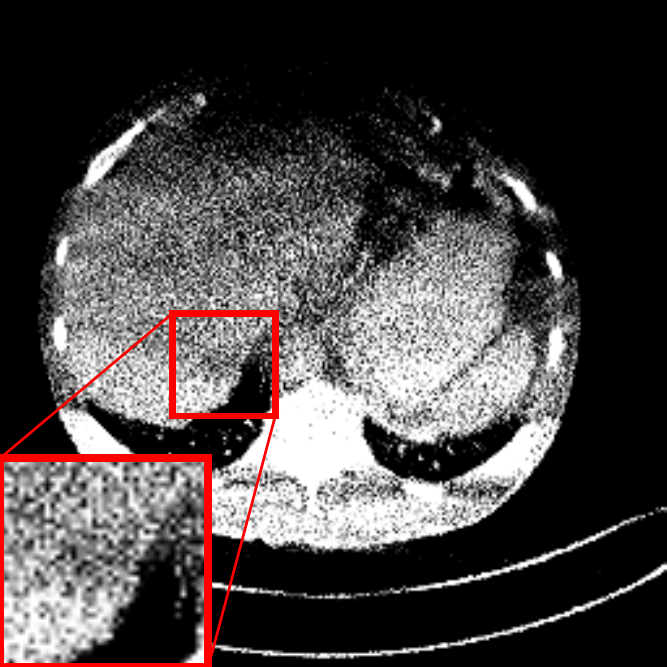}
}
\end{minipage}
\begin{minipage}{0.16\linewidth}
\subfigure[ 78\,HU]{
\includegraphics[width = \linewidth]{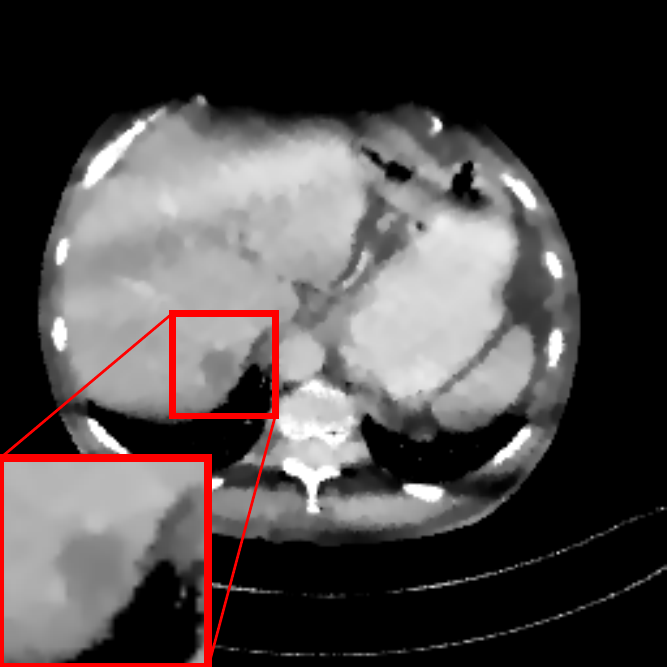}
}
\end{minipage}
\begin{minipage}{0.16\linewidth}
\subfigure[ 68\,HU]{
\includegraphics[width = \linewidth]{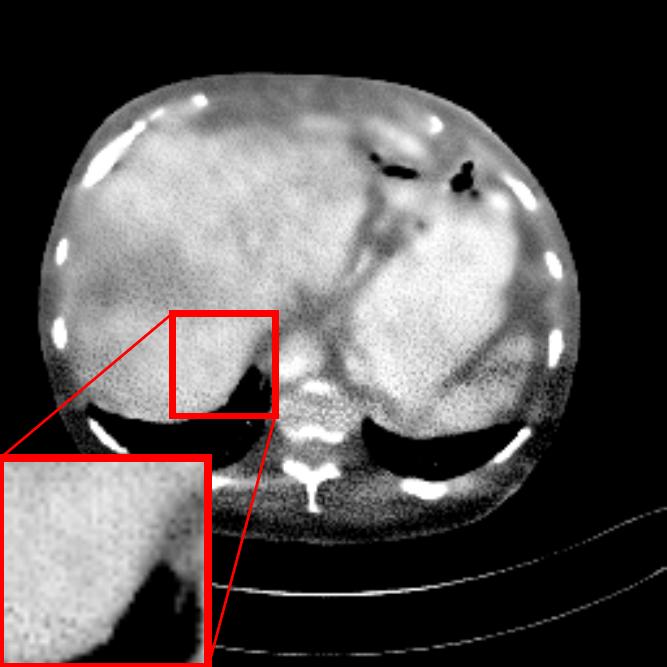}
}
\end{minipage}
\begin{minipage}{0.16\linewidth}
\subfigure[50\,HU]{
\includegraphics[width = \linewidth]{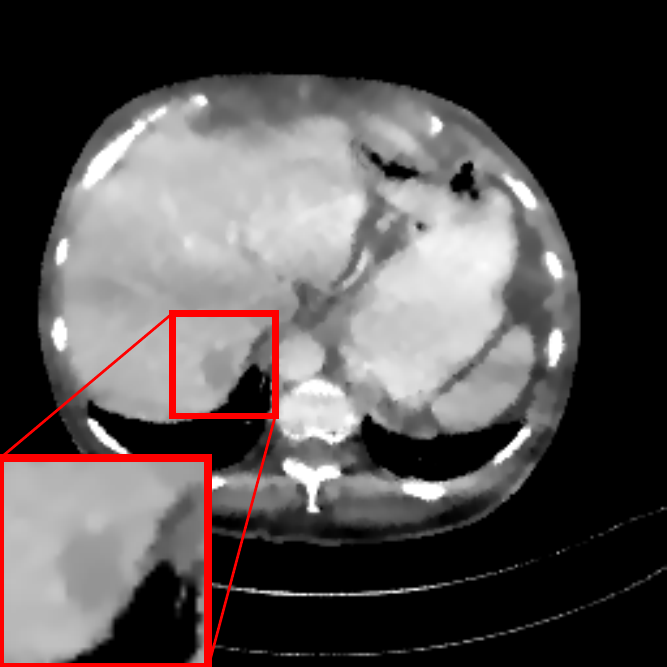}
}
\end{minipage}
\caption{Results of one example slice (the 196th slice of Patient NO.\,3) in $150^\circ$ cone-beam limited-angle tomography with Poisson noise displayed in window [-1000, 1000]\,HU (top row) and in window [-200, 200]\,HU (bottom row). The lesion in the ROI of (i) reconstructed by the U-Net disappears entirely, while it is well reconstructed by DCR in (j). The SSIM index and the RMSE value for each method are displayed in the top row and the bottom row respectively.}
\label{Fig:LimitedAngleResults}
\end{figure*}

The reconstruction results of one example slice (the 196th slice of Patient NO.\,3) in $150^\circ$ cone-beam limited-angle tomography with Poisson noise are displayed in Fig.\,\ref{Fig:LimitedAngleResults}. The top row images are displayed in a wide window of [-1000, 1000]\,HU. The top body outline is severely distorted due to missing data in the FBP reconstruction in Fig.\,\ref{Fig:LimitedAngleResults}(b). Meanwhile, it also suffers from Poisson noise. In the wTV reconstruction in Fig.\,\ref{Fig:LimitedAngleResults}(c), the Poisson noise is reduced. Due to the large angular range of missing data, wTV is able to restore the top body outline only partially. In Fig.\,\ref{Fig:LimitedAngleResults}(d), the body outline is fully restored by the U-Net. However, not all structures are accurate in the U-Net reconstruction, when it is redisplayed in a narrow window of [-200, 200]\,HU in Fig.\,\ref{Fig:LimitedAngleResults}(i). Especially, the lesion in the ROI is hardly seen, while it is well observed in the ROIs of the reference image (Fig.\,\ref{Fig:LimitedAngleResults}(f)) and the wTV reconstruction (Fig.\,\ref{Fig:LimitedAngleResults}(h)). By combining deep learning with compressed sensing, the lesion is reconstructed again in the DCR reconstruction in Fig.\,\ref{Fig:LimitedAngleResults}(j), while the body outline is reconstructed as well. 
\subsection{Sparse-View CT}

\begin{figure*}[tbh]
\centering
\begin{minipage}{0.16\linewidth}
\centering
$\vf_{\text{reference}}$
\end{minipage}
\begin{minipage}{0.16\linewidth}
\centering
$\vf_{\text{FBP}}$
\end{minipage}
\begin{minipage}{0.16\linewidth}
\centering
$\vf_{\text{wTV}}$
\end{minipage}
\begin{minipage}{0.16\linewidth}
\centering
$\vf_{\text{U-Net}}$
\end{minipage}
\begin{minipage}{0.16\linewidth}
\centering
$\vf_{\text{DCR}}$
\end{minipage}

\vspace{3pt}

\begin{minipage}{0.16\linewidth}
\subfigure[]{
\includegraphics[width = \linewidth]{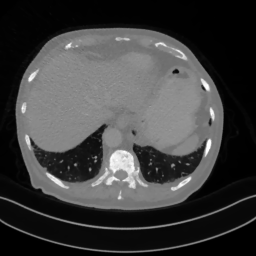}
}
\end{minipage}
\begin{minipage}{0.16\linewidth}
\subfigure[ 0.988]{
\includegraphics[width = \linewidth]{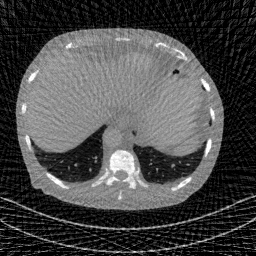}
}
\end{minipage}
\begin{minipage}{0.16\linewidth}
\subfigure[ 0.998]{
\includegraphics[width = \linewidth]{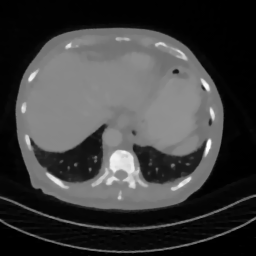}
}
\end{minipage}
\begin{minipage}{0.16\linewidth}
\subfigure[ 0.990]{
\includegraphics[width = \linewidth]{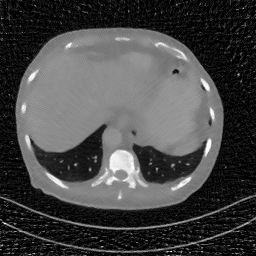}
}
\end{minipage}
\begin{minipage}{0.16\linewidth}
\subfigure[0.999]{
\includegraphics[width = \linewidth]{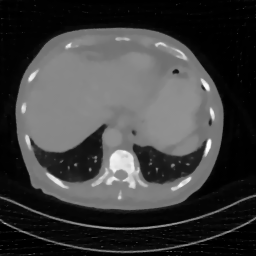}
}
\end{minipage}

\begin{minipage}{0.16\linewidth}
\subfigure[]{
\includegraphics[width = \linewidth]{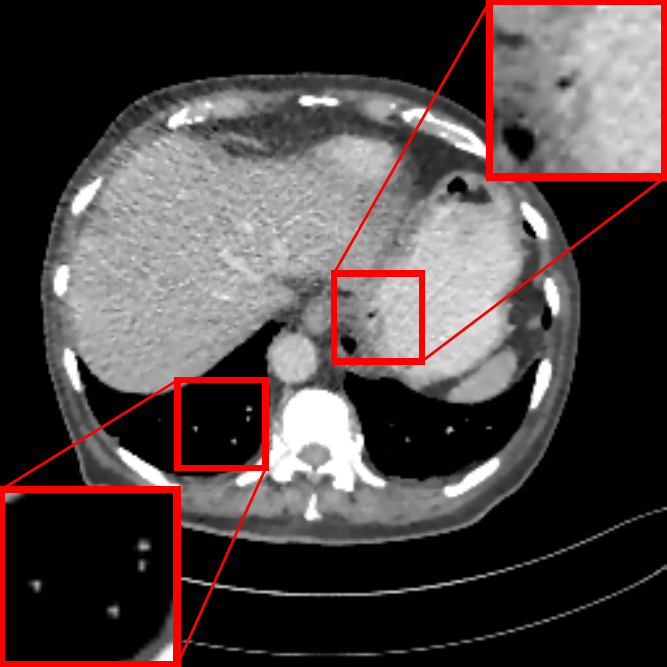}
}
\end{minipage}
\begin{minipage}{0.16\linewidth}
\subfigure[ 104\,HU]{
\includegraphics[width = \linewidth]{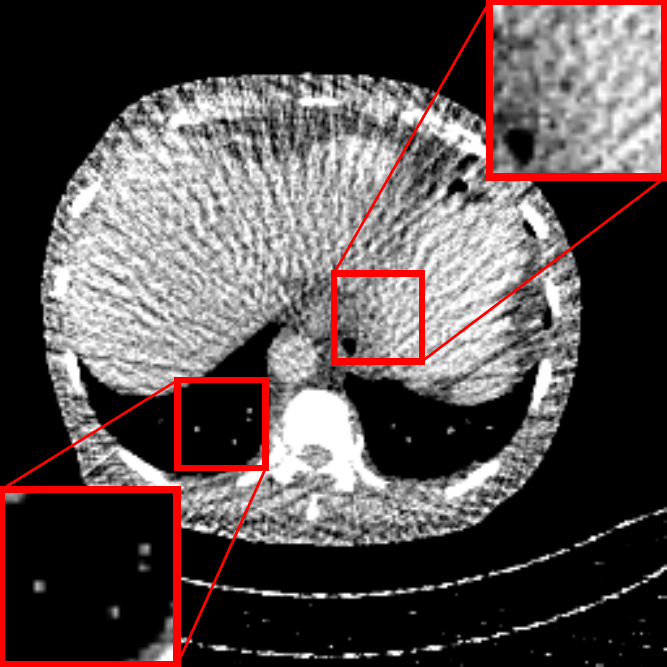}
}
\end{minipage}
\begin{minipage}{0.16\linewidth}
\subfigure[ 35\,HU]{
\includegraphics[width = \linewidth]{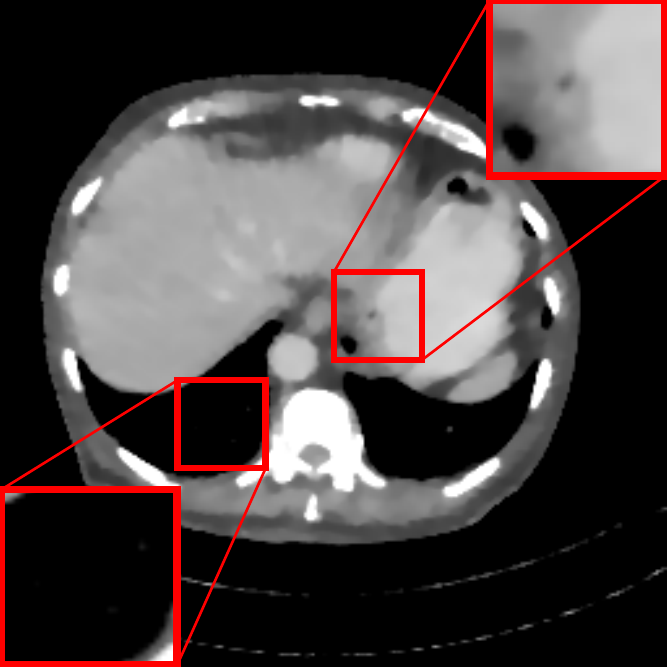}
}
\end{minipage}
\begin{minipage}{0.16\linewidth}
\subfigure[ 50\,HU]{
\includegraphics[width = \linewidth]{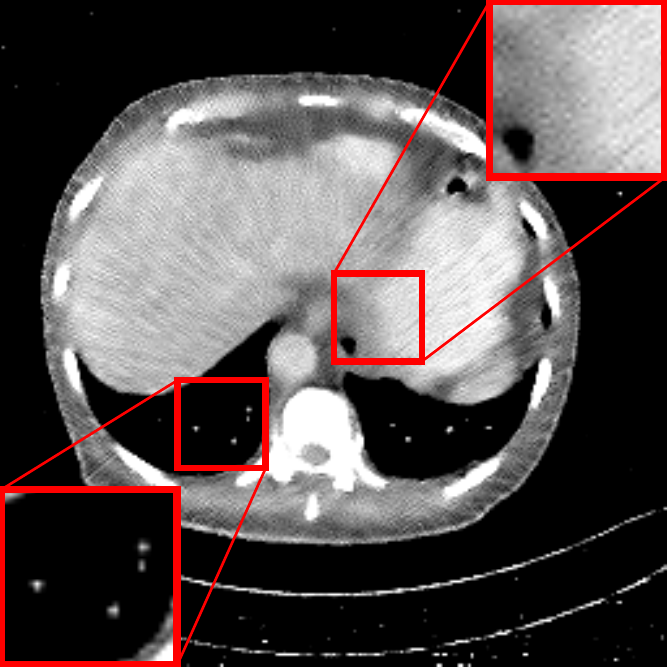}
}
\end{minipage}
\begin{minipage}{0.16\linewidth}
\subfigure[34\,HU ]{
\includegraphics[width = \linewidth]{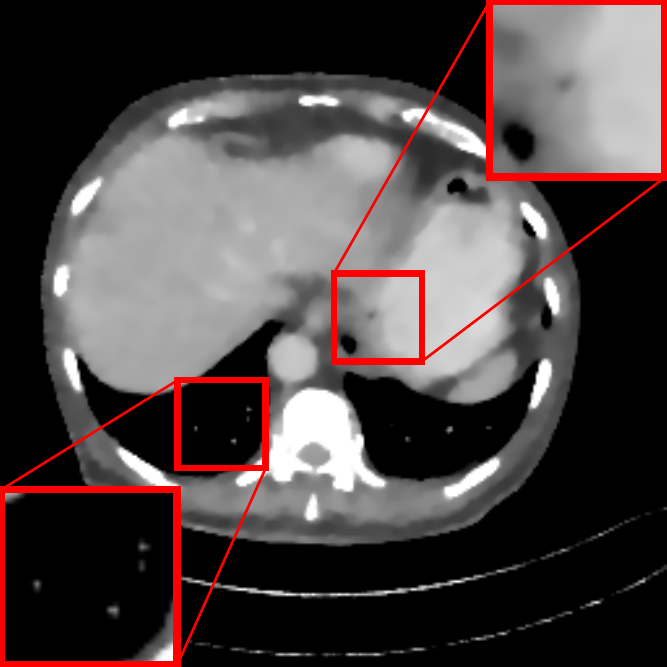}
}
\end{minipage}

\caption{Results of one example slice (the 162th slice of Patient NO.\,3) in sparse-view (90 projections) cone-beam CT displayed in window [-1000, 1000]\,HU (top row) and in window [-200, 200]\,HU (bottom row). The lung vessels in the bottom left ROI of (h) reconstructed by wTV are not visible in this window. The dark cavity at the center of the top right ROI of (i) is not reconstructed by the U-Net. The SSIM index and the RMSE value for each method are displayed in the top row and the bottom row respectively.}
\label{Fig:SparseViewResults}
\end{figure*}

The reconstruction results of one slice (the 162th slice of Patient NO.\,3) in sparse-view (90 projections) cone-beam CT are displayed in Fig.\,\ref{Fig:SparseViewResults}. The top row images are displayed in a wide window of [-1000, 1000]\,HU. In the FBP reconstruction in Fig.\,\ref{Fig:SparseViewResults}(b), streak artifacts and aliasing are observed. The high frequency aliasing and streaks are reduced effectively by wTV, as displayed in Fig.\,\ref{Fig:SparseViewResults}(c). In the U-Net reconstruction displayed in Fig.\,\ref{Fig:SparseViewResults}(d), the streak artifacts and aliasing are reduced to some degree. However, artifacts remain in the background area and the patient bed area. The streak artiacts and aliasing are reduced effectively in the DCR reconstruction in Fig.\,\ref{Fig:SparseViewResults}(e).
In the narrow window, most anatomical structures are obscured by the streak artifacts and aliasing in the FBP reconstruction in Fig.\,\ref{Fig:SparseViewResults}(g). For wTV in Fig.\,\ref{Fig:SparseViewResults}(h), most anatomical structures are reconstructed. However, the intensities of the lung vessels in the bottom left ROI, which consist of a few pixels, are reduced. As a consequence, they are not visible in the given window in Fig.\,\ref{Fig:SparseViewResults}(h). Note that they are still visible in the wide window in Fig.\,\ref{Fig:SparseViewResults}(c). For the U-Net reconstruction in Fig.\,\ref{Fig:SparseViewResults}(i), the dark cavity structure in the center of the top right ROI is not reconstructed. Nevertheless, the lung vessels in the bottom left ROI are preserved. In the DCR reconstruction in Fig.\,\ref{Fig:SparseViewResults}(j), both the lung vessels and the dark cavity structure in the two ROIs are reconstructed.

\section{Discussion}

Deep learning based methods achieve encouraging image reconstructions from insufficient data, as displayed in Figs.\,\ref{Fig:TruncationNoiseFreeLW}(e) and (k), Fig.\,\ref{Fig:LimitedAngleResults}(d), and Fig.\,\ref{Fig:SparseViewResults}(d). In a wide display window, high contrast structures (e.g., bones) and body outlines are typically reconstructed with high confidence. However, many fine details maybe reconstructed incorrectly when they are displayed in a narrow window, although they appear very realistic. In this work, false negative lesion cases of deep learning methods are exhibited in Fig.\,\ref{Fig:truncationCorrectionNoisyCase}(k), Fig.\,\ref{Fig:LimitedAngleResults}(i), and Fig.\,\ref{Fig:SparseViewResults}(d). Moreover, false positive lesion cases of deep learning are discovered, as shown in Figs.\,\ref{Fig:FakeLesions}(e) and (k). Especially, the lesion generated by the U-Net in Fig.\,\ref{Fig:FakeLesions}(e) is so realistic that radiology experts may draw false diagnostic conclusions. Therefore, the observations in this work serve as a warning to the robustness of deep learning in clinical applications. 

Insufficient training data and noise are two main potential factors to influence the robustness of deep learning, as indicated by the experiments in this work. In the noise-free scenarios, false positive lesions are observed in the U-Net results in Figs.\,\ref{Fig:FakeLesions}(e) and (k). This is potentially caused by the insufficient training data, as in this work only 425 slices are used for training in each experiment due to the limited access to patient data. In addition, the observations in Figs.\,\ref{Fig:truncationCorrectionNoisyCase}(e) and (k), Fig.\,\ref{Fig:LimitedAngleResults}(i), and Fig.\,\ref{Fig:SparseViewResults}(i) demonstrate that deep learning is very sensitive to noise. In deep neural networks, as a consequence of high dimensional dot products, noise will accumulate layer by layer and adds up to a large change to the output \cite{goodfellow2014explaining}. Therefore, even if noise has a small magnitude, it still has a severe impact on the output images. In the noisy cases, the U-Net is trained on data with Poisson noise, which endows the U-Net the ability to deal with noisy images to some degree, as we observed previously \cite{huang2018some}. However, it is still not sufficient to reduce Poisson noise entirely as indicated by Fig.\,\ref{Fig:truncationCorrectionNoisyCase}(k), or it tends to over-smooth images as indicated by Fig.\,\ref{Fig:truncationCorrectionNoisyCase}(e). In either way, fine structures are lost. 

Instability is a general problem for deep learning in solving inverse problems \cite{gottschling2020troublesome}. Therefore, generating reconstructed images directly from a neural network appears inadequate. Our proposed DCR method, with the help of compressed sensing, has superior performance than deep learning only for different CT reconstruction scenarios with insufficient data. For truncated data, Fig.\,\ref{Fig:TruncationNoiseFreeSW} demonstrates that DCR can improve the image quality of learning-based images both inside and outside the FOV. Fig.\,\ref{Fig:FakeLesions} and Fig.\,\ref{Fig:truncationCorrectionNoisyCase} show that DCR can correct false negative and false positive lesions in learning-based images, which highlights the important clinical value of DCR. For limited-angle data and sparse-view data, our example results in Fig.\,\ref{Fig:LimitedAngleResults} and Fig.\,\ref{Fig:SparseViewResults} demonstrate the efficacy of DCR as well.

As a hybrid method combining deep learning with compressed sensing, DCR preserves their respective advantages: a) DCR is as robust as compressed sensing, since both of them keep the data consistency with measured data. b) DCR is more efficient than compressed sensing. For DCR, 10 iterations only (Fig.\,\ref{Fig:iterationNumber}) are sufficient to converge using the learned images as initialization, while compressed sensing only typically requires much more iterations. c) DCR is more effective for image reconstruction from insufficient data. Compressed sensing only typically fails to reconstruct the regions where a large amount of data are missing, as indicated by Figs.\,\ref{Fig:TruncationNoiseFreeLW}(d) and (j), Fig.\,\ref{Fig:LimitedAngleResults}(c), and Fig.\,\ref{Fig:SparseViewResults}(h). Thanks to the information provided by the learned image prior for the missing data, DCR is able to reconstruct those regions better than compressed sensing only.

\section{Conclusion}
In this work, the robustness of deep learning for CT image reconstruction from insufficient data is investigated. Particularly, false positive and false negative lesion cases generated by the state-of-the-art U-Net are exemplified for image reconstruction from truncated data, limited-angle data and sparse-view data. To improve deep learning reconstruction, the DCR method is proposed combining the advantages of deep learning and compressed sensing. It utilizes compressed sensing to compute a final reconstruction which is consistent with the measured data, while using the learned reconstruction as prior for unmeasured data. In such a combination, the high representation power of deep learning and the high robustness of compressed sensing are integrated together in the proposed DCR method.


%

%

\ 

\textbf{Disclaimer:} The concepts and information presented in this paper are based on research and are not commercially available.

\ifCLASSOPTIONcaptionsoff
  \newpage
\fi

\end{document}